\documentclass[fleqn,usenatbib]{mnras}

\usepackage{newtxtext,newtxmath}
 
\usepackage[T1]{fontenc}

\DeclareRobustCommand{\VAN}[3]{#2}
\let\VANthebibliography\thebibliography
\def\thebibliography{\DeclareRobustCommand{\VAN}[3]{##3}\VANthebibliography}

\usepackage{graphicx}
\usepackage{amsmath}
\usepackage{amssymb}
\usepackage{longtable}
\usepackage{subfig}
\title[Emission lines at $z\sim 2.3$]{The MOSDEF Survey: A Comprehensive Analysis of the Rest-optical Emission-line Properties of $\MakeLowercase{z}\sim 2.3$ Star-forming Galaxies$^{1}$}

\author[J. N. Runco et al.]{Jordan N. Runco,$^{2}$\thanks{E-mail: jrunco@astro.ucla.edu}
Alice E. Shapley,$^{2}$
Ryan L. Sanders,$^{3,4}$
Michael W. Topping,$^{2}$\newauthor
Mariska Kriek,$^{5}$
Naveen A. Reddy,$^{6}$
Alison L. Coil,$^{7}$
Bahram Mobasher,$^{6}$\newauthor
Brian Siana,$^{6}$
William R. Freeman,$^{6}$
Irene Shivaei,$^{8,4}$
Mojegan Azadi,$^{9}$\newauthor
Sedona H. Price,$^{10}$
Gene C. K. Leung,$^{7}$
Tara Fetherolf,$^{6}$
Laura de Groot,$^{11}$\newauthor
Tom Zick,$^{5}$
Francesca M. Fornasini,$^{9}$
Guillermo Barro$^{12}$
\\
$^{1}$Based on data obtained at the W.M. Keck Observatory, which is operated as a scientific partnership among the California Institute of \\ Technology, the University of California,  and the National Aeronautics and Space Administration, and was made possible by the generous  \\ financial support  of the W.M. Keck Foundation.\\
$^{2}$Physics \& Astronomy Department, University of California: Los Angeles, 430 Portola Plaza, Los Angeles, CA 90095, USA\\
$^{3}$Department of Physics, University of California, Davis, One Shields Ave, Davis, CA 95616, USA\\
$^{4}$Hubble Fellow\\
$^{5}$Astronomy Department, University of California, Berkeley, CA 94720, USA\\
$^{6}$Department of Physics \& Astronomy, University of California, Riverside, 900 University Avenue, Riverside, CA 92521, USA\\
$^{7}$Center for Astrophysics and Space Sciences, University of California, San Diego, 9500 Gilman Dr., La Jolla, CA 92093-0424, USA\\
$^{8}$Department of Astronomy/Steward Observatory, 933 North Cherry Ave, Rm N204, Tucson, AZ 85721-0065, USA\\
$^{9}$Harvard-Smithsonian Center for Astrophysics, 60 Garden Street, Cambridge, MA, 02138, USA \\
$^{10}$Max-Planck-Institut f\"ur Extraterrestrische Physik, Postfach 1312, Garching, 85741, Germany \\
$^{11}$Department of Physics, The College of Wooster, 1189 Beall Avenue, Wooster, OH 44691, USA\\
$^{12}$Department of Phyics, University of the Pacific, 3601 Pacific Ave, Stockton, CA 95211, USA
}

\date{Accepted XXX. Received YYY; in original form ZZZ}

\pubyear{2021}

\begin{document}
\label{firstpage}
\pagerange{\pageref{firstpage}--\pageref{lastpage}}
\maketitle

\begin{abstract}
We analyze the rest-optical emission-line spectra of $z\sim 2.3$ star-forming galaxies in the complete MOSFIRE Deep Evolution Field (MOSDEF) survey.  In investigating the origin of the well-known offset between the sequences of high-redshift and local galaxies in the [O~\textsc{III}]$\lambda$5008/H$\beta$ vs. [N~\textsc{II}]$\lambda$6585/H$\alpha$ (``[N~\textsc{II}] BPT'') diagram, we define two populations of $z\sim 2.3$ MOSDEF galaxies.  These include the \textit{high} population that is offset towards higher [O~\textsc{III}]$\lambda$5008/H$\beta$ and/or [N~\textsc{II}]$\lambda$6585/H$\alpha$ with respect to the local SDSS sequence and the \textit{low} population that overlaps the SDSS sequence.  These two groups are also segregated within the [O~\textsc{III}]$\lambda$5008/H$\beta$ vs. [S~\textsc{II}]$\lambda\lambda$6718,6733/H$\alpha$ and the [O~\textsc{III}]$\lambda\lambda$4960,5008/[O~\textsc{II}]$\lambda\lambda$3727,3730 (O$_{32}$) vs. ([O~\textsc{III}]$\lambda\lambda$4960,5008+[O~\textsc{II}]$\lambda\lambda$3727,3730)/H$\beta$ (R$_{23}$) diagrams, which suggests qualitatively that star-forming regions in the more offset galaxies are characterized by harder ionizing spectra at fixed nebular oxygen abundance.  We also investigate many galaxy properties of the split sample and find that the \textit{high} sample is on average smaller in size and less massive, but has higher specific star-formation rate and star-formation-rate surface density values and is slightly younger compared to the \textit{low} population. From Cloudy+BPASS photoionization models, we estimate that the \textit{high} population has a lower stellar metallicity (i.e., harder ionizing spectrum) but slightly higher nebular metallicity and higher ionization parameter compared to the \textit{low} population.  While the \textit{high} population is more $\alpha$-enhanced (i.e., higher $\alpha$/Fe) than the \textit{low} population, both samples are significantly more $\alpha$-enhanced compared to local star-forming galaxies with similar rest-optical line ratios. These differences must be accounted for in {\it all} high-redshift star-forming galaxies -- not only those ``offset'' from local excitation sequences.
\end{abstract}

\begin{keywords}
galaxies: evolution --- galaxies: high-redshift -- galaxies: ISM
\end{keywords}

\section{Introduction} \label{sec:intro}

Rest-frame optical emission-line spectroscopy is one of the most important observational tools for understanding galaxy formation and evolution.  Such spectra provide a wealth of information about a galaxy, including its star-formation rate (SFR), dust extinction, active galactic nucleus (AGN) activity, virial and non-virial dynamics (e.g., outflows) and properties of the ionized interstellar medium such as the metallicity, electron density ($n_{\rm{e}}$), and ionization parameter ($U$; i.e., the ratio of ionizing photon density to hydrogen, and therefore electron, density).  Accordingly, knowledge about the emission-line properties of star-forming galaxies across cosmic time is essential for understanding the evolution of the stellar and gaseous content in galaxies.  It is of particular interest to study rest-optical emission-line spectra of galaxies at $z\sim 2$, which represents the epoch of peak star formation in the universe \citep{mad14} and a time before the modern Hubble sequence was fully in place.

In early work, \citet{bal81} showed how diagnostic diagrams measuring the intensity ratios of [O~\textsc{III}]$\lambda$5008/H$\beta$ vs. [N~\textsc{II}]$\lambda$6585/H$\alpha$, i.e., the \textquotedblleft [N~\textsc{II}] BPT diagram,'' can be used to distinguish between star formation and AGN activity as the ionizing source for a galaxy.  Other authors have also considered the [O~\textsc{III}]$\lambda$5008/H$\beta$ vs. [S~\textsc{II}]$\lambda\lambda$6718,6733/H$\alpha$ diagram (originally introduced in \citealt{vei87} and referred to hereafter as the \textquotedblleft [S~\textsc{II}] BPT diagram'') when investigating the ionization mechanism of a galaxy.  Indeed, the shape of the ionizing spectrum as well as the typical ionization parameter are different in gas excited by an AGN as opposed to by hot stars. Accordingly, AGNs and star-forming galaxies occupy distinct regions within rest-optical emission-line diagrams.  These diagnostic diagrams also reveal information about the physical properties of the galaxies.  For example, stellar mass ($M_{\ast}$) and metallicity have been found to increase with decreasing [O~\textsc{III}]$\lambda$5008/H$\beta$ and increasing [N~\textsc{II}]$\lambda$6585/H$\alpha$ along the local star-forming sequence in the [N~\textsc{II}] BPT diagram (e.g., \citealt{mas16}).

Another diagram commonly used to describe star-forming galaxies is the [O~\textsc{III}]$\lambda\lambda$4960,5008/[O~\textsc{II}]$\lambda\lambda$3727,3730 (O$_{32}$) vs. ([O~\textsc{III}]$\lambda\lambda$4960,5008+[O~\textsc{II}]$\lambda\lambda$3727,3730)/H$\beta$ (R$_{23}$) diagram.  O$_{32}$ and R$_{23}$ are rough tracers for ionization parameter and metallicity, respectively.  This diagram allows us to probe such physical quantities in star-forming galaxies (e.g. \citealt{lil03, nak13}).  For local galaxies, there is an increase in metallicity from the high excitation end (high O$_{32}$ \& R$_{23}$) to the low excitation tail (low O$_{32}$ \& R$_{23}$) on this diagram \citep{and13, sha15}.

Early studies with Keck/NIRSPEC found that the location of galaxies on the [N~\textsc{II}] BPT diagram is redshift dependent, as galaxies with $z>1$ are found to be offset from local Sloan Digital Sky Survey (SDSS; \citealt{yor00}) galaxies, showing elevated [O~\textsc{III}]$\lambda$5008/H$\beta$ at fixed [N~\textsc{II}]$\lambda$6585/H$\alpha$ (or vice versa; e.g. \citealt{sha05, erb06, liu08}).  Based on new observations with multi-object near-infrared spectrographs on 8-10-meter class telescopes, the sample of high-redshift galaxies with measurements of the BPT diagram emission lines now numbers in the hundreds.  Of note, two large surveys of the high-redshift BPT diagram include the MOSFIRE Deep Evolution Field (MOSDEF; \citealt{kri15}) survey and the Keck Baryonic Structure Survey (KBSS: \citealt{ste14}), which find, based on much more robust statistical evidence, that high-redshift galaxies tend to have elevated [O~\textsc{III}]$\lambda$5008/H$\beta$ and/or [N~\textsc{II}]$\lambda$6585/H$\alpha$ compared to local galaxies \citep{ste14, sha15, sha19, san16, str17}.  

Understanding this offset in emission-line ratios is vital because we use strong rest-optical emission-lines as empirical tracers for many physical properties (e.g., gas-phase oxygen abundance \citealt{pet04}) of the interstellar medium (ISM).  Due to the observed offset for $z>1$ galaxies, it is unclear if local metallicity calibrations, e.g., \citet{pet04}, yield accurate metallicities when applied in the high-redshift universe.  Therefore, it is essential to gain a complete understanding of why high-redshift galaxies have elevated [O~\textsc{III}]$\lambda$5008/H$\beta$ and/or [N~\textsc{II}]$\lambda$6585/H$\alpha$ compared to the local SDSS sample.  

There have been many proposed explanations of this systematic offset.  Possible explanations include variations in physical properties of galaxies such as H~\textsc{II} region electron densities (or proportionally pressures), density structure, H~\textsc{II} region ionization parameter, H~\textsc{II} region ionizing spectra at fixed metallicities, gas-phase N/O abundance ratio differences, unresolved AGN activity, and shocks (e.g. \citealt{liu08, bri08, wri10, kew13, yeh13, mas14, coi15, san16, ste16, str17, fre19, kas19, sha19, top20a}).  Galaxy selection effects could also be potential factors in this offset \citep{jun14}.  

Preliminary results from the MOSDEF survey \citep{sha15, san16} suggested that the offset in emission-line ratios is primarily due to elevated N/O at fixed O/H abundance patterns in $z\sim 2.3$ galaxies relative to local ones.  Studies using KBSS data \citep{ste16, str17, str18} argued for a harder ionizing spectrum at fixed nebular metallicity as the main cause of the offset while other works using the Fiber Multi-Object Spectrograph (FMOS)-COSMOS survey \citep{kas17, kas19} or local analogues of high redshift galaxies \citep{bia20} attribute the observed [N~\textsc{II}] BPT offset to a higher ionization parameter.  
More recent results from the MOSDEF survey using the complete MOSDEF data set now suggest a harder ionizing spectrum drives the BPT offset \citep{sha19, san20a, san20b, top20a}.

The completed MOSDEF survey provides an ideal data set of high-redshift galaxies to help explore the observed offset on the [N~\textsc{II}] BPT diagram.  This survey provides access to spectra of $\sim$1500 galaxies in the $z\sim 1.4-3.8$ redshift range, with full spectroscopic coverage of all the emission-lines needed to complete the [N~\textsc{II}] BPT, [S~\textsc{II}] BPT, and O$_{32}$ vs. R$_{23}$ diagrams in the $z\sim 1.4-2.6$ redshift range.  

In this work, we improve upon the previous $z>1$ BPT offset studies based on early MOSDEF data by \citet{sha15} and \citet{san16}.  We now have the full MOSDEF sample in hand and apply a more careful spectral energy distribution (SED) fitting method that incorporates corrections to broadband photometric measurements for rest-optical emission-line fluxes. Accordingly, we derive unbiased age and stellar mass measurements, and can conduct a more thorough and complete investigation of the emission-line ratio properties of the MOSDEF sample.  As an example of the level of improvement enabled by the full MOSDEF sample, the initial work by, \citet{sha15} explored the location of 53 star-forming galaxies at $1.9\leq z\leq 2.7$ on the [N~\textsc{II}] BPT diagram, and investigated the $M_{\ast}$, specific SFR (sSFR), and SFR surface density ($\Sigma_{\rm{SFR}}$) of the sample.  In this study, the sample increases to 180 star-forming galaxies on the [N~\textsc{II}] BPT diagram, and we also investigate additional galaxy parameters such as the galaxy effective radius ($R_{\rm{e}}$), SFR, $n_{\rm{e}}$, and stellar population age.  \citet{san16} investigated how a sample of 53 star-forming galaxies at $1.9\leq z\leq 2.7$ translated from the [N~\textsc{II}] BPT diagram to the [S~\textsc{II}] BPT diagram and O$_{32}$ vs. R$_{23}$ diagram.  In the current study, that number is more than doubled to 122.  Our more comprehensive analysis enables a better understanding of the observed offset between the MOSDEF sample and local galaxies on the [N~\textsc{II}] BPT diagram.  Through the analysis outlined above, we aim to determine the underlying physical causes of this offset.

The paper is organized as follows.  In Section \ref{sec:methods}, we briefly describe the MOSDEF survey, and review our survey sample selection and data reduction.  Section \ref{sec:results} presents the results of this study.  Section \ref{sec:discussion} discusses the results and how they relate to photoionization models.  Section \ref{sec:summary} presents a summary of key results and looks ahead to future analyses.  We adopt the following abbreviations for emission-line ratios used frequently throughout the paper.
\begin{equation}
\label{eqn:N2_abbreviation}
    \rm{N2 = [N~\textsc{II}]\lambda6585/H\alpha}
\end{equation}
\begin{equation}
\label{eqn:S2_abbreviation}
    \rm{S2 = [S~\textsc{II}]\lambda\lambda6718,6733/H\alpha}
\end{equation}
\begin{equation}
\label{eqn:O3_abbreviation}
    \rm{O3 = [O~\textsc{III}]\lambda5008/H\beta}
\end{equation}
\begin{equation}
\label{eqn:R23_abbreviation}
    \rm{R_{23} = ([O~\textsc{III}]\lambda\lambda4960,5008+[O~\textsc{II}]\lambda\lambda3727,3730)/H\beta}
\end{equation}
\begin{equation}
\label{eqn:O32_abbreviation}
    \rm{O_{32} = [O~\textsc{III}]\lambda\lambda4960,5008/[O~\textsc{II}]\lambda\lambda3727,3730}
\end{equation}
\begin{equation}
\label{eqn:O3N2_abbreviation}
    \rm{O3N2 = O3/N2}
\end{equation}
O$_{32}$ and R$_{23}$ have already been defined, but we include them here for completeness.  All emission-line wavelengths are vacuum wavelengths.  Throughout this paper, we adopt a $\Lambda$-CDM cosmology with $H_0$ = 70 km s$^{-1}$ Mpc$^{-1}$, $\Omega_m$ = 0.3, and $\Omega_\Lambda$ = 0.7.  Also, we assume the solar abundance pattern from \citet{asp09}.

\section{Methods} \label{sec:methods}

\subsection{MOSDEF Sample \& Ancillary Measurements} \label{subsec:mosdef_sample}

The MOSDEF survey was a 48.5-night observing program spanning multiple years (2012-2016), using the MultiObject Spectrometer For Infra-Red Exploration (MOSFIRE; \citealt{mcl12}) on the 10 m Keck I telescope telescope.  The survey targets three redshift ranges: $1.37\leq z\leq 1.70$, $2.09\leq z\leq 2.61$, and $2.95\leq z\leq 3.80$.  These ranges are chosen to optimize the detection of strong rest-optical emission lines within windows of atmospheric transmission.  There are 354, 678, and 266 galaxies that have been spectroscopically confirmed in the MOSDEF low, medium, and high-redshift ranges, respectively.  The sample is H-band magnitude limited, and is located in the well-studied CANDELS and 3D-\textit{HST} legacy fields \citep{gro11, koe11, mom16}: AEGIS, COSMOS, GOODS-N, GOODS-S, and UDS.  The ancillary data from these fields enables us to probe other properties of the sample (e.g., galaxy size and stellar mass).

\begin{figure}
    \centering
    \includegraphics[scale=0.5]{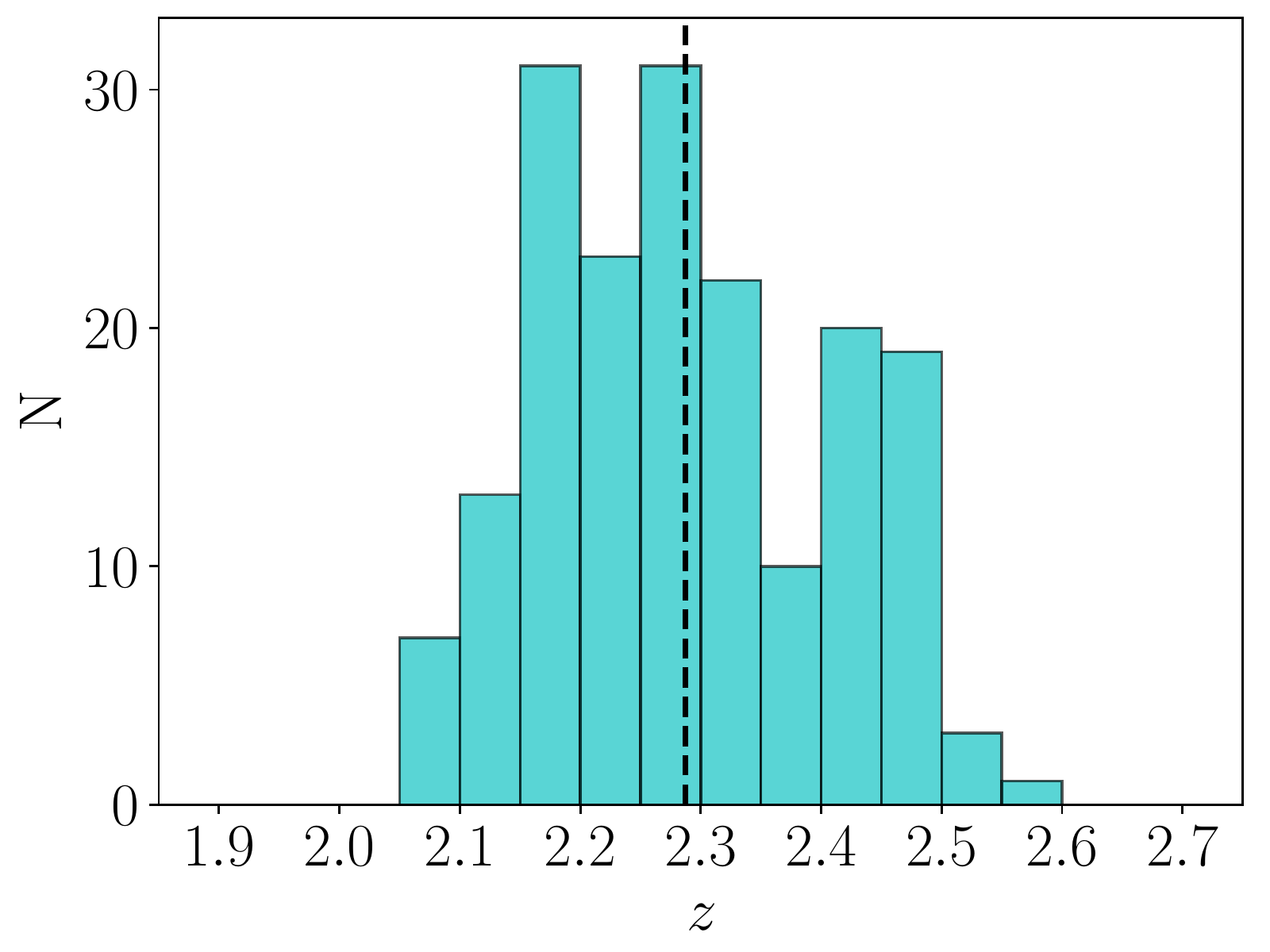}
    \caption{Redshift distribution of the 180 star-forming MOSDEF galaxies at $z\sim 2.3$ with [N~\textsc{II}] BPT diagram classifications (i.e., S/N $\geq$ 3 detections in H$\beta$, [O~\textsc{III}]$\lambda\lambda$4960,5008, H$\alpha$, and [N~\textsc{II}]$\lambda$6585).  The dashed vertical line displays the median redshift of the sample, $z_{\rm med}=2.29$.}
    \label{fig:redshift_histogram}
\end{figure}

In this study, we restrict the sample to the central redshift bin, spanning the redshift range to $1.9\leq z\leq 2.7$ to reflect the scatter between target photometric and measured spectroscopic redshifts (i.e., the fact that not all spectroscopic redshifts, when actually measured, fell precisely within the target $2.09\leq z\leq 2.61$ window).  For galaxies in this redshift range, [O~\textsc{II}]$\lambda\lambda$3727,3730, H$\beta$, [O~\textsc{III}]$\lambda\lambda$4960,5008, H$\alpha$, [N~\textsc{II}]$\lambda$6585, and [S~\textsc{II}]$\lambda\lambda$6718,6733 features are captured within the J, H, and K near-IR filter wavelength ranges.  For MOSFIRE, the J, H, and K filters have typical respective wavelength coverages of 1.142-1.365 $\mu$m, 1.450-1.826 $\mu$m, and 1.897-2.427 $\mu$m and spectral resolutions of R = 3000, 3650, and 3600.  

We imposed further restrictions to reach our final sample.  Galaxies containing an AGN were identified and removed from the sample based on their IR colors, X-ray luminosity, or if N2 $>$ 0.5 \citep{coi15, aza17, leu17}.  We also require a S/N $\geq$ 3 for each emission-line in our analysis.  These criteria yielded a sample of 180 galaxies for initial classification in the [N~\textsc{II}] BPT diagram, of which 122 galaxies are additionally detected in [S~\textsc{II}]$\lambda\lambda$6718,6733 and [O~\textsc{II}]$\lambda\lambda$3727,3730 and have size measurements from the \textit{Hubble Space Telescope} (\textit{HST}). The redshift distribution of the 180 MOSDEF star-forming galaxies at $z\sim 2.3$ with [N~\textsc{II}] BPT diagram classifications is shown in Figure \ref{fig:redshift_histogram}.

We correct H$\alpha$ and H$\beta$ line fluxes for stellar Balmer absorption as described in \citet{kri15} and \citet{red15}.  We also dust correct emission-line ratios for which the member features differ significantly in wavelength, including O$_{32}$ and R$_{23}$.  For such corrections, we assumed the \citet{car89} dust attenuation curve and an unreddened H$\alpha$/H$\beta$ ratio of 2.86.  In the case of emission-line ratios for which the lines are close in wavelength (O3, N2, and S2) dust correction was not applied.  

In this study, we investigate additional galaxy photometric, spectroscopic, and structural properties for the MOSDEF sample in order to try to better understand the observed rest-optical emission-line properties.  The \citet{hao11} calibration for a \citet{cha03} initial mass function (IMF) and solar metallicity is used to estimate SFR(H$\alpha$) from stellar-Balmer-absorption-corrected, dust-corrected, and slit-loss-corrected H$\alpha$ luminosities \citep{red15, kri15, shi15}.  We used the SED fitting code, FAST \citep{kri09} to obtain key stellar population parameters including stellar mass and age. For this modeling, as a default we assumed star-formation histories of the delayed-$\tau$ form, where SFR $\propto t \times e^{-t/\tau}$. Here $t$ is the time since the onset of star formation (i.e., age), and $\tau$ is the characteristic star-formation timescale. Given the range of best-fit $\tau$ values, the meaning of absolute ages is not necessarily clear. In order to obtain a better gauge of the relative maturities of the galaxy stellar populations in our sample, we used both normalized ages (i.e., $t/\tau$), and also the age ($t$) obtained from constant star formation (CSF) models.  Galaxy sizes, $R_{\rm{e}}$, are taken as the F160W galaxy half-light radii from the \citet{van14} catalog, which were estimated using single-component S\'ersic profile fits to the two-dimensional light distribution of galaxies in the CANDELS and 3D-\textit{HST} fields.  We combine SFR(H$\alpha$) and $R_{\rm{e}}$ to estimate the SFR surface denisty, $\Sigma_{\rm{SFR}}$ as:
\begin{equation}
\label{eqn:sfr_surface_density}
    \Sigma_{\rm{SFR}} = \frac{\rm{SFR(H\alpha)}}{2\pi R_{\rm{e}}^{2}}
\end{equation}
We estimate the sSFR using SFR(H$\alpha$) and $M_{\ast}$, both of which are described above.  

The electron density, $n_{\rm{e}}$, is estimated with the [O~\textsc{II}] and [S~\textsc{II}] emission-line doublets using the method described in \citet{san16}.  In that study, the [O~\textsc{II}] and [S~\textsc{II}] electron densities agree within the uncertainties; both reliable tracers of the density within H~\textsc{II} regions.  [O~\textsc{II}] is the preferred choice because the doublet typically has a higher S/N.  We only use [S~\textsc{II}] measurements when we do not have adequate [O~\textsc{II}] data.  This scenario arises for the following reasons:
\begin{enumerate}
    \item Low S/N.  A S/N $\geq$ 3 is required for each emission-line in the doublet to estimate $n_{\rm{e}}$; however, a less restrictive requirement of S/N $\geq$ 3 for the overall doublet is sufficient for a galaxy to be included in the [S~\textsc{II}] BPT and O$_{32}$ vs. R$_{23}$ diagrams.  We only require S/N $\geq$ 3 for the individual doublet members when investigating the $n_{\rm{e}}$ of the sample.
    \item Rejection of the [O~\textsc{II}] doublet based on visual inspection.  Such cases occur when skylines affect the [O~\textsc{II}] doublet but not the [S~\textsc{II}] doublet, spurious detections, and poor fits.
\end{enumerate}

We set the lower limit for $n_{\rm{e}}$ to be 1 cm$^{-3}$.

\begin{figure*}
    \centering
    \includegraphics[scale=0.55]{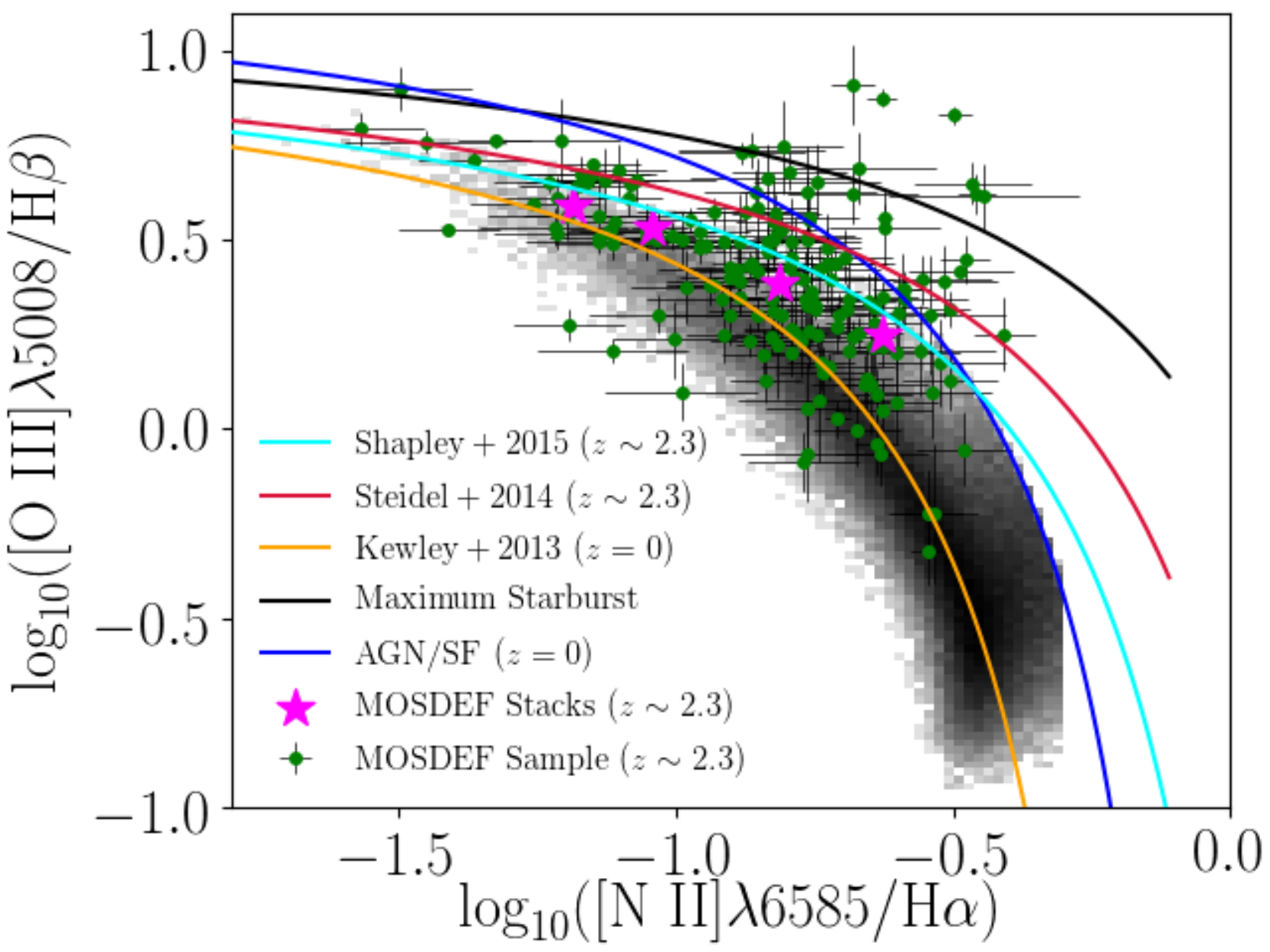}
    \caption{[N~\textsc{II}] BPT diagram.  Green points indicate $1.9\leq z\leq 2.7$ MOSDEF galaxies with S/N $\geq$ 3 for all four plotted emission-lines.  Large magenta stars represent measurements of stacks from composite spectra, binned by stellar mass, for $1.9\leq z\leq 2.7$ MOSDEF galaxies.  The grayscale 2D histogram indicates local SDSS galaxies.  The cyan curve is the fit to the MOSDEF sample from \citet{sha15}.  The orange curve is a fit to the $z\sim 0$ star-forming locus \citep{kew13}.  The red curve is the best fit to the $z\sim 2.3$ galaxies from the Keck Baryonic Structure Survey (KBSS; \citealt{ste14}).  The black curve is the maximum starburst line from \citet{kew01}.  The blue curve is the empirical AGN/star-forming galaxy dividing line from \citet{kau03}.} \label{fig:nii_bpt_diagram}
\end{figure*}

\subsection{SDSS Comparison Sample} \label{subsec:sdss_sample}

Throughout this study, we compare our high-redshift MOSDEF sample to local galaxies.  For this comparison, we use archival data from the SDSS Data Release 7 (DR7; \citealt{aba09}).  We obtain galaxy properties and emission-line measurements from the MPA-JHU DR7 release of spectrum measurements\footnote[13]{https://wwwmpa.mpa-garching.mpg.de/SDSS/DR7/}.  SDSS galaxies are selected within the $0.04\leq z\leq 0.10$ redshift range.  We impose similar restrictions on our SDSS sample to those applied to the MOSDEF sample by requiring that each emission-line used in the analysis has a S/N $\geq$ 3.  For the SDSS sample, we remove AGN using equation 1 from \citet{kau03}.  Galaxies are also identified as having an AGN component if N2 $>$ 0.5.  These criteria result in a comparison sample of 103,422 SDSS galaxies when considering the [N~\textsc{II}] BPT diagram alone, and 74,726 SDSS galaxies when considering galaxies with simultaneous detections across all three emission-line diagrams analyzed in this work ([N~\textsc{II}] BPT, [S~\textsc{II}] BPT, and O$_{32}$ vs. R$_{23}$).

\section{Results}\label{sec:results}

\subsection{The [N~\textsc{II}] BPT Diagram} \label{subsec:the_niibpt_diagram}

We start by investigating the locations of $z\sim 2.3$ star-forming galaxies in the [N~\textsc{II}] BPT diagram based on the complete MOSDEF sample (Figure \ref{fig:nii_bpt_diagram}).  We include the 180 galaxies with $\geq 3\sigma$ detections for all four emission-lines (H$\beta$, [O~\textsc{III}]$\lambda\lambda$4960,5008, H$\alpha$, and [N~\textsc{II}]$\lambda$6585) and the corresponding SDSS sample with the same four emission-lines detected.  Similar to previous MOSDEF studies (e.g. \citealt{sha15, san16, sha19, top20a}) and other studies from the literature (e.g., \citealt{sha05, erb06, ste14}), there is a systematic offset observed for the high-redshift MOSDEF sample from the local sequence (Figure \ref{fig:nii_bpt_diagram}).  The MOSDEF galaxies appear on average to be shifted towards the AGN region of the diagram with elevated N2 and/or O3 values, with some galaxies on the AGN side of the \citet{kau03} AGN/SF boundary.  There are even a small number of galaxies past the maximum starburst line from \citet{kew01} as well.

To search for any biases in the sample based on the $\geq 3\sigma$ detection requirement in all four [N~\textsc{II}] BPT lines, we construct spectral stacks for all MOSDEF galaxies at $1.9\leq z\leq 2.7$ with H$\alpha$ emission detected at S/N $\geq$ 3. There are four spectral stacks, divided into bins of stellar mass (see \citealt{san18} for a full description of the method for constructing composite spectra).  The emission-line ratios measured from stacked spectra follow the distribution of datapoints measured from MOSDEF galaxies with individual detections in all [N~\textsc{II}] BPT lines.  Therefore, the sample of individual detections seems to represent the parent $z\sim 2.3$ MOSDEF data set with minimal bias.  The four stacks, similarly to the individually detected galaxies, are offset from the local SDSS sample with elevated N2 and/or O3 values.

\begin{figure*}
    \centering
    \includegraphics[width=0.49\linewidth]{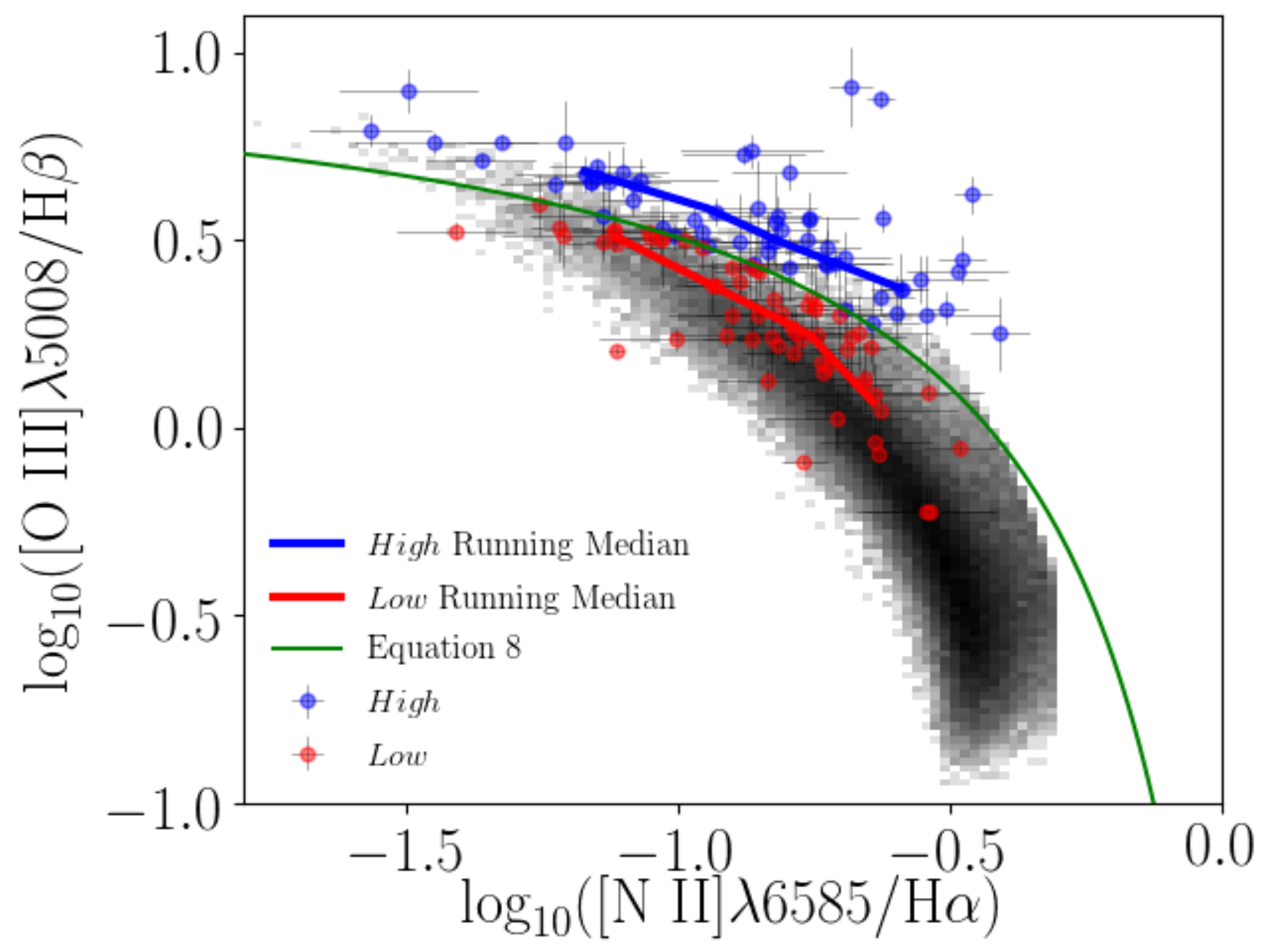}
    \includegraphics[width=0.49\linewidth]{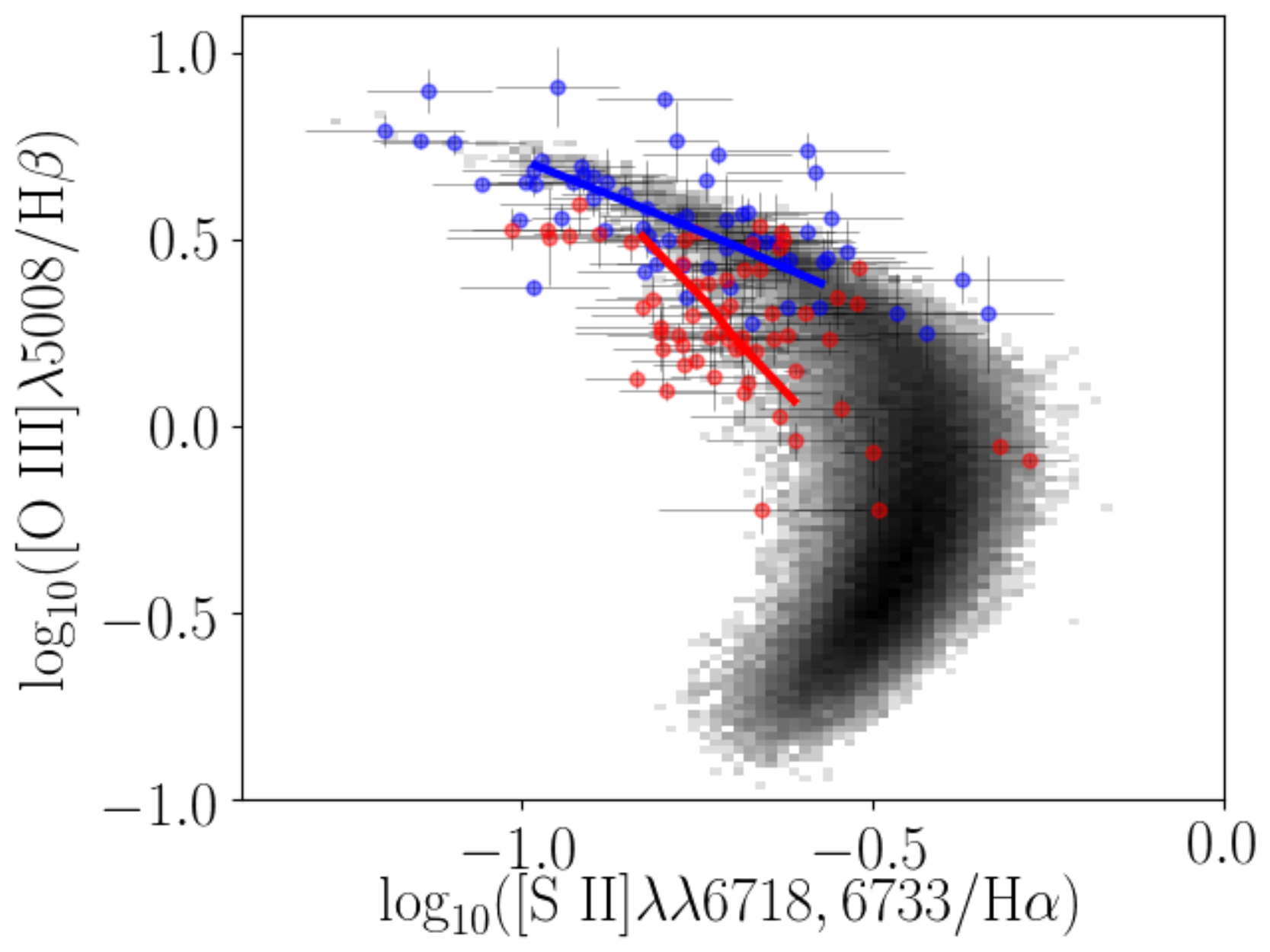}
    \includegraphics[width=0.49\linewidth]{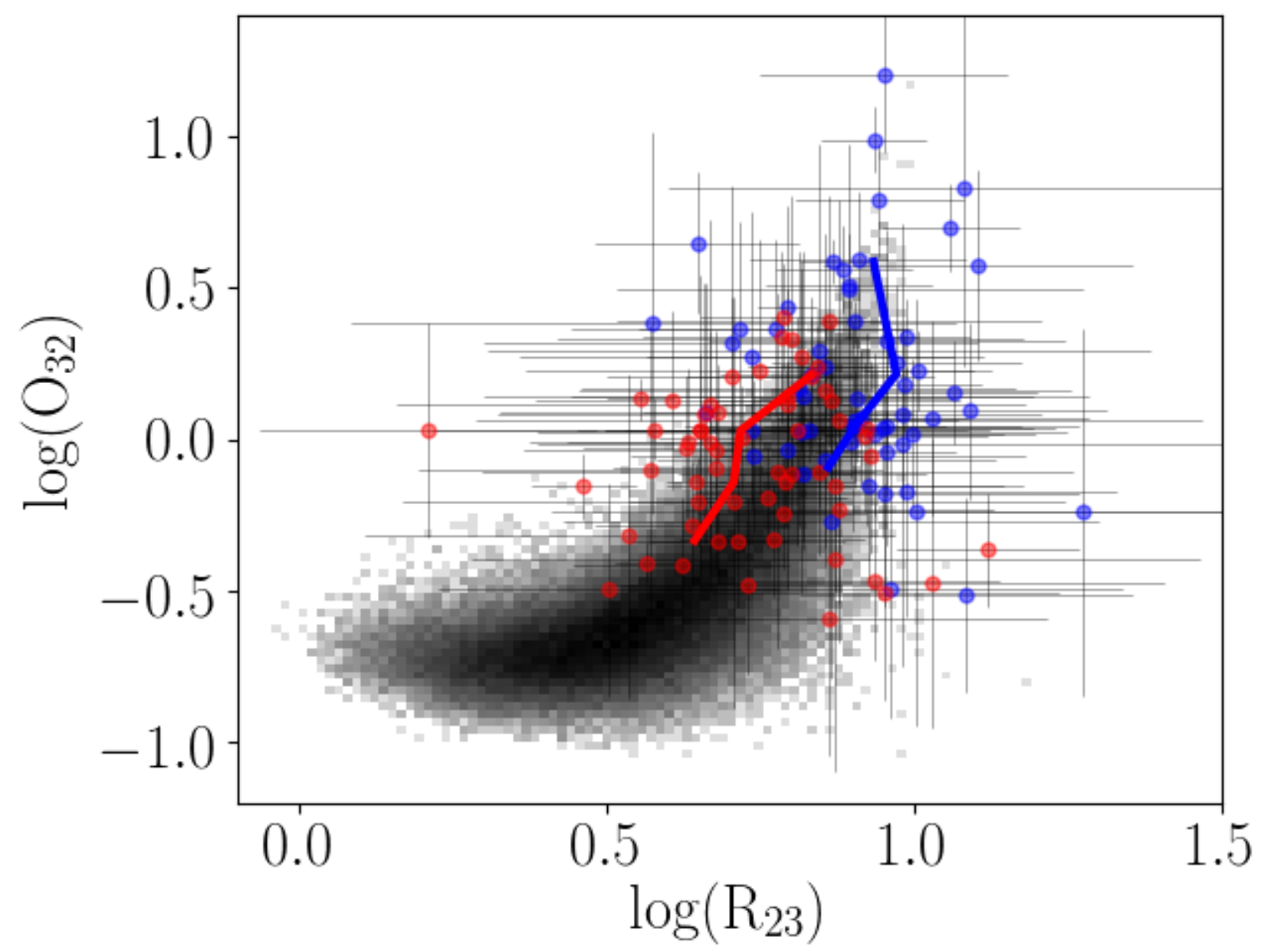}
    \caption{The sample of 122 MOSDEF galaxies that have $\geq 3\sigma$ detections in [O~\textsc{II}]$\lambda\lambda$3727,3730, H$\beta$, [O~\textsc{III}]$\lambda\lambda$4960,5008, H$\alpha$, [N~\textsc{II}]$\lambda$6585, and [S~\textsc{II}]$\lambda\lambda$6718,6733.  Top left: we modify the MOSDEF fit from \citet{sha15} by altering the y-intercept, so a curve with the same functional form splits the data set evenly into two populations.  The red galaxies (\textit{low} sample) fall close to the local SDSS sequence, while the blue galaxies (\textit{high} sample) are shifted off the local sequence.  Top right: The same two groups of data points as in the left panel.  The red and blue lines indicate binned medians as described in Section \ref{subsec:nii_bpt_split}.  These lines show that the two groups are also separated in the [S~\textsc{II}] BPT diagram, with the \textit{high} sample higher in S2 and/or O3 than the \textit{low} sample.  Bottom center: The same two groups of data points as in the top panels.  The red and blue lines again indicate binned medians as described in Section \ref{subsec:nii_bpt_split}.  The two groups are segregated in the O$_{32}$ vs. R$_{23}$ diagram as well, with the \textit{high} sample having a higher median R$_{23}$ values at fixed O$_{32}$ where there is overlap in O$_{32}$.} \label{fig:nii_bpt_split}
\end{figure*}

\subsection{Dividing the $z\sim 2.3$ Sample in the [N~\textsc{II}] BPT Diagram} \label{subsec:nii_bpt_split}

Figure \ref{fig:nii_bpt_diagram} shows that on the [N~\textsc{II}] BPT diagram, part of the MOSDEF sample sits on or near the local SDSS sequence, while the remainder of the galaxies lie off the SDSS sequence and are shifted towards the local AGN region.  We now investigate if the location of a high-redshift galaxy $-$ either on or offset from the local SDSS sequence $-$ is connected with other physical properties of the galaxy.  For this analysis, we examine how galaxies on the [N~\textsc{II}] BPT diagram populate the [S~\textsc{II}] BPT and O$_{32}$ vs. R$_{23}$ diagrams, and therefore require the sample to have $\geq 3\sigma$ detections for all emission-lines on these diagrams: [O~\textsc{II}]$\lambda\lambda$3727,3730, H$\beta$, [O~\textsc{III}]$\lambda\lambda$4960,5008, H$\alpha$, [N~\textsc{II}]$\lambda$6585, and [S~\textsc{II}]$\lambda\lambda$6718,6733.  There are 123 MOSDEF galaxies that meet this criterion.  One galaxy was removed because it does not have a $R_{\rm{e}}$ measurement, which brings the final sample to 122 MOSDEF galaxies.

For this analysis, we split the final sample of 122 galaxies into two groups using the functional form presented in \citet{sha15}, and adjusting the y-intercept so that our sample is divided in half.  The equation used to split the MOSDEF sample into two groups of 61 galaxies each is
\begin{equation}
\label{eqn:NII_BPT_split}
    \textup{log}\big(\textup{[O~\textsc{III}]/H}\beta\big) = \frac{0.67}{\textup{log}(\textup{[N~\textsc{II}]/H}\alpha) - 0.20} + 1.065
\end{equation} 
The new curve is 0.055 dex lower in [O~\textsc{III}]$\lambda$5008/H$\beta$ compared to equation 1 in \citet{sha15}.  This splitting of the sample on the [N~\textsc{II}] BPT diagram is shown in the upper left panel of Figure \ref{fig:nii_bpt_split}, where galaxies above the curve are indicated with blue symbols and those below with red symbols.  Hereafter, we refer to the galaxies above the curve on the [N~\textsc{II}] BPT diagram as the \textit{high} sample, and the group below the curve as the \textit{low} sample, following the nomenclature in \citet{top20a}.  We plot the two samples on both the [S~\textsc{II}] BPT diagram (upper right panel of Figure \ref{fig:nii_bpt_split}) and the O$_{32}$ vs. R$_{23}$ diagram (bottom panel of Figure \ref{fig:nii_bpt_split}).  We also include binned median lines for both the \textit{high} (blue) and \textit{low} (red) populations.  These binned medians are binned by log$_{10}$(O3N2) for the [N~\textsc{II}] BPT diagram, log$_{10}$(O3) $-$ log$_{10}$(S2) for the [S~\textsc{II}] BPT diagram, and log$_{10}$(O$_{32}$) $+$ log$_{10}$(R$_{23}$) for the O$_{32}$ vs. R$_{23}$ diagram.  The binning schemes above were adopted because they divide our sample into subgroups segregated roughly along the local star-forming sequence.  For both the \textit{high} and \textit{low} samples, each of which contain 61 galaxies, there are four equally sized bins (three bins of 15 galaxies and one bin of 16).

For completeness, we checked for potential bias between the smaller sample of 122 galaxies with detections across all three emission-line diagrams and the parent sample of 180 galaxies defined based on detections in the [N~\textsc{II}] BPT diagram alone.  Of the 58 galaxies that were removed, 32 are above the curve dividing our sample (Equation \ref{eqn:NII_BPT_split}) and 26 are below it on the [N~\textsc{II}] BPT diagram.  Because these two groups are approximately equal in size, we can expect that the smaller subset with [O~\textsc{II}] and [S~\textsc{II}] also detected (122 galaxies) has approximately the same average BPT offset as the larger [N~\textsc{II}] BPT only sample (180 galaxies).  Therefore, we can conclude that the cuts to create the sample of 122 galaxies with detections across all three emission-line diagrams do not introduce biases when compared to the larger parent sample of 180 galaxies.

\subsubsection{Division in the Other Diagrams} \label{subsec:sii_bpt_o32r23_splits}

We find separation between the \textit{high} and \textit{low} samples on both the [S~\textsc{II}] BPT and O$_{32}$ vs. R$_{23}$ diagrams.  For the [S~\textsc{II}] BPT diagram, the \textit{high} sample has a systematically higher S2 at fixed O3 (or vice versa).  For the O$_{32}$ vs. R$_{23}$ diagram, the \textit{high} sample has a systematically higher R$_{23}$ value at fixed O$_{32}$.  

Our new results on the [S~\textsc{II}] BPT and O$_{32}$ vs. R$_{23}$ diagrams update the early MOSDEF results from \citet{san16}.  In this earlier work, based on a significantly smaller sample of only 53 galaxies, subsamples split in the [N~\textsc{II}] BPT diagram were found to be well mixed in the [S~\textsc{II}] BPT and O$_{32}$ vs. R$_{23}$ diagrams.  In contrast, our results are consistent with those of \citet{str17}, based on the KBSS survey. In that study, the sample was split into two groups based on [N~\textsc{II}] BPT diagram location $-$ large and small offset from the local SDSS sequence.  These subsamples were found to remain separated on the [S~\textsc{II}] BPT diagram.  \citet{str17} did not investigate if this segregation remained on the O$_{32}$ vs. R$_{23}$ diagram.  Section \ref{sec:discussion} contains a more in-depth comparison with the results from \citet{str17} and a discussion of the implications for the underlying causes of the observed shift in the high-redshift [N~\textsc{II}] BPT diagram.

\begin{figure*}
    \centering
     \subfloat[]{
       \includegraphics[width=0.32\linewidth]{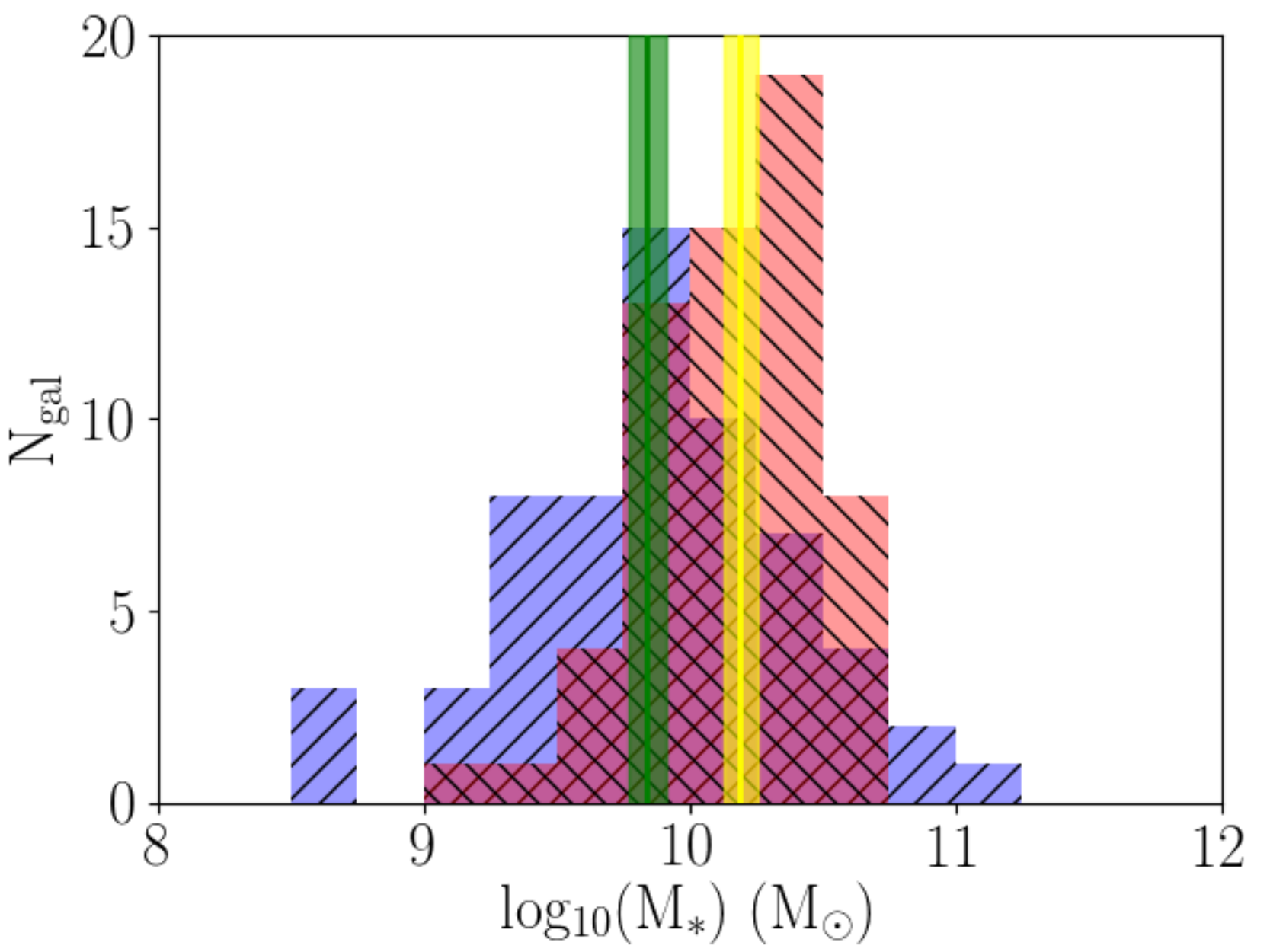}
     }
     \hfill
     \subfloat[]{
       \includegraphics[width=0.32\linewidth]{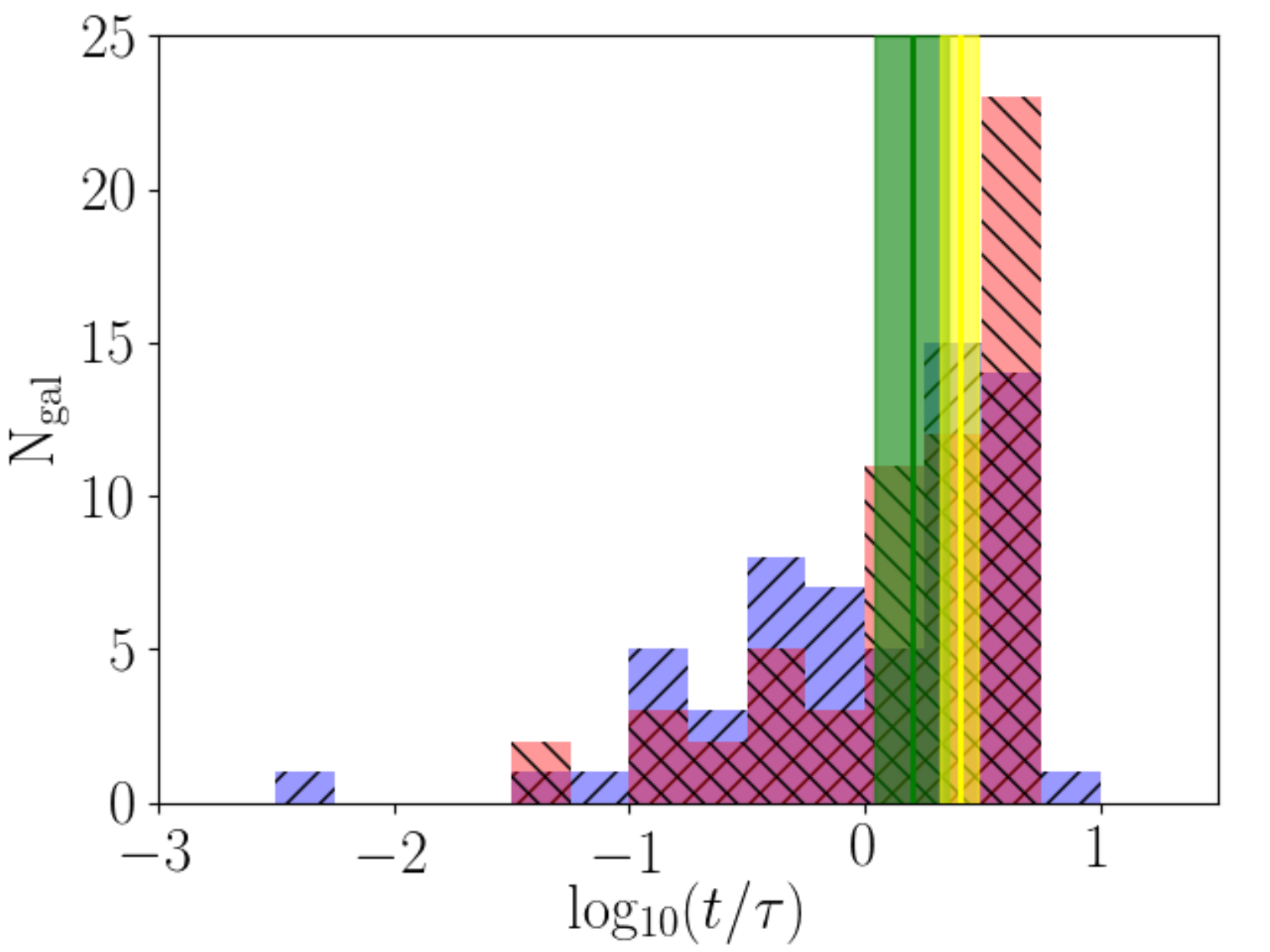}
     }
     \hfill
     \subfloat[]{
       \includegraphics[width=0.32\linewidth]{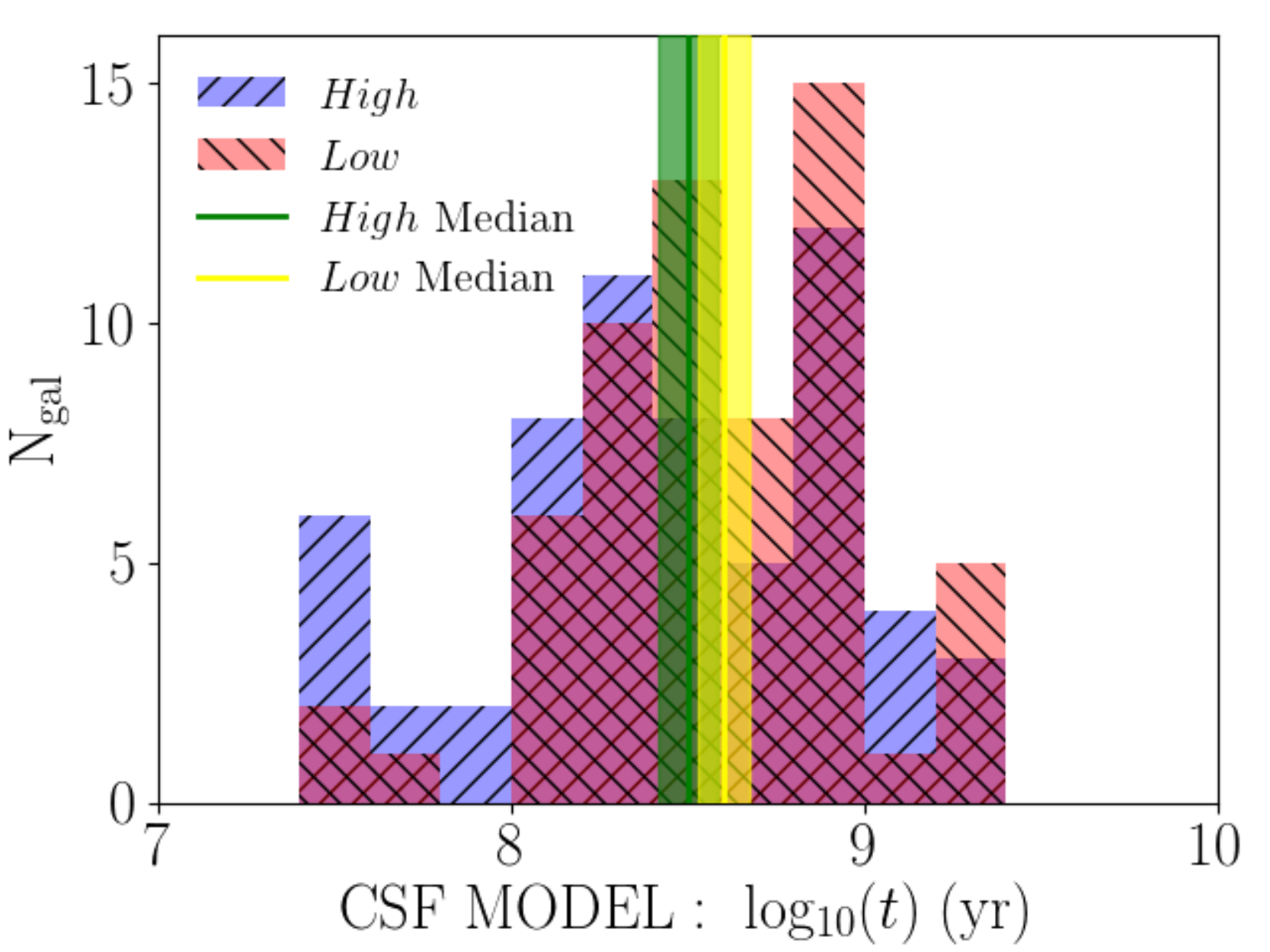}
     }
     \hfill
     \subfloat[]{
       \includegraphics[width=0.32\linewidth]{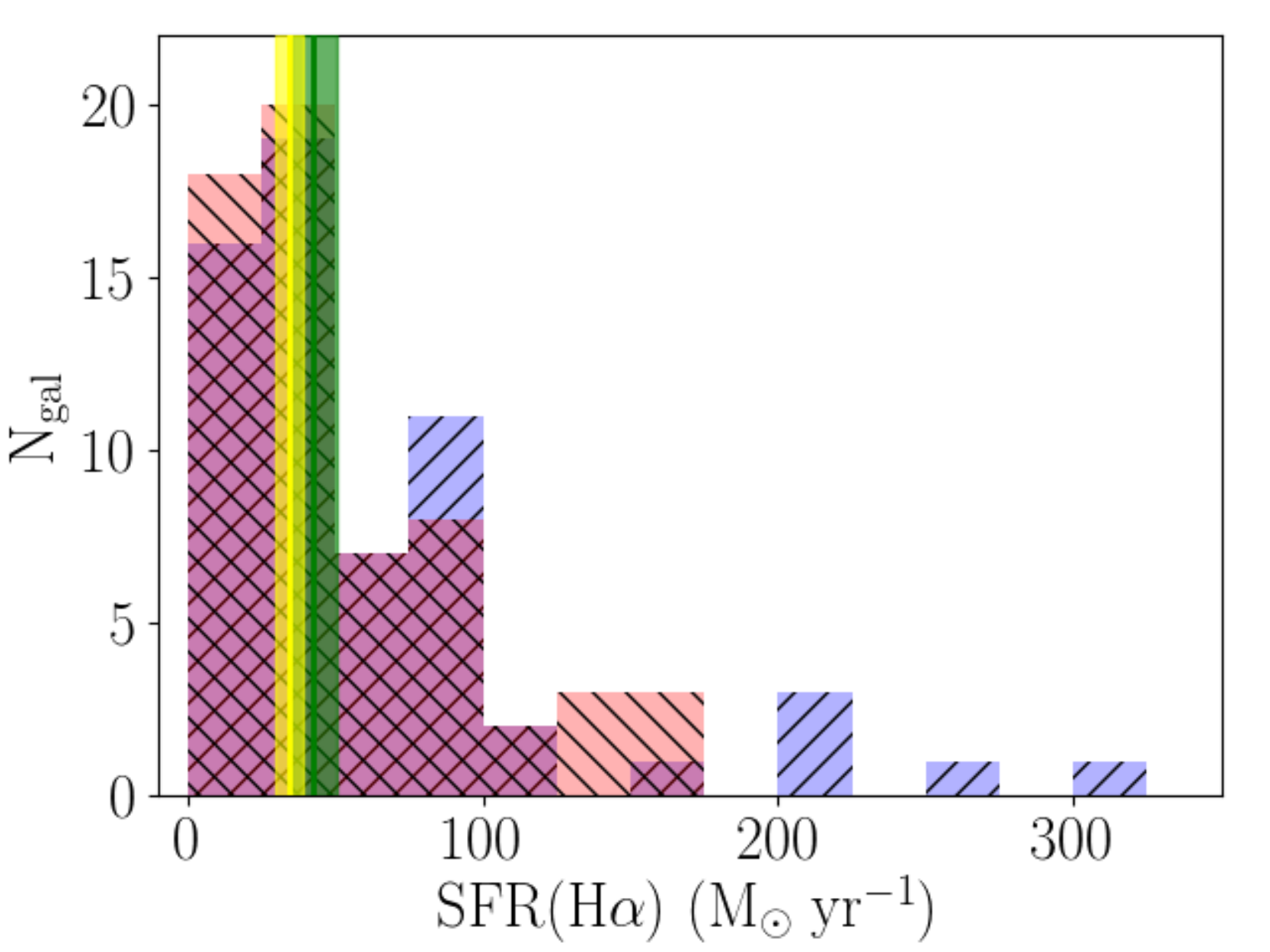}
     }
     \hfill
     \subfloat[]{
       \includegraphics[width=0.32\linewidth]{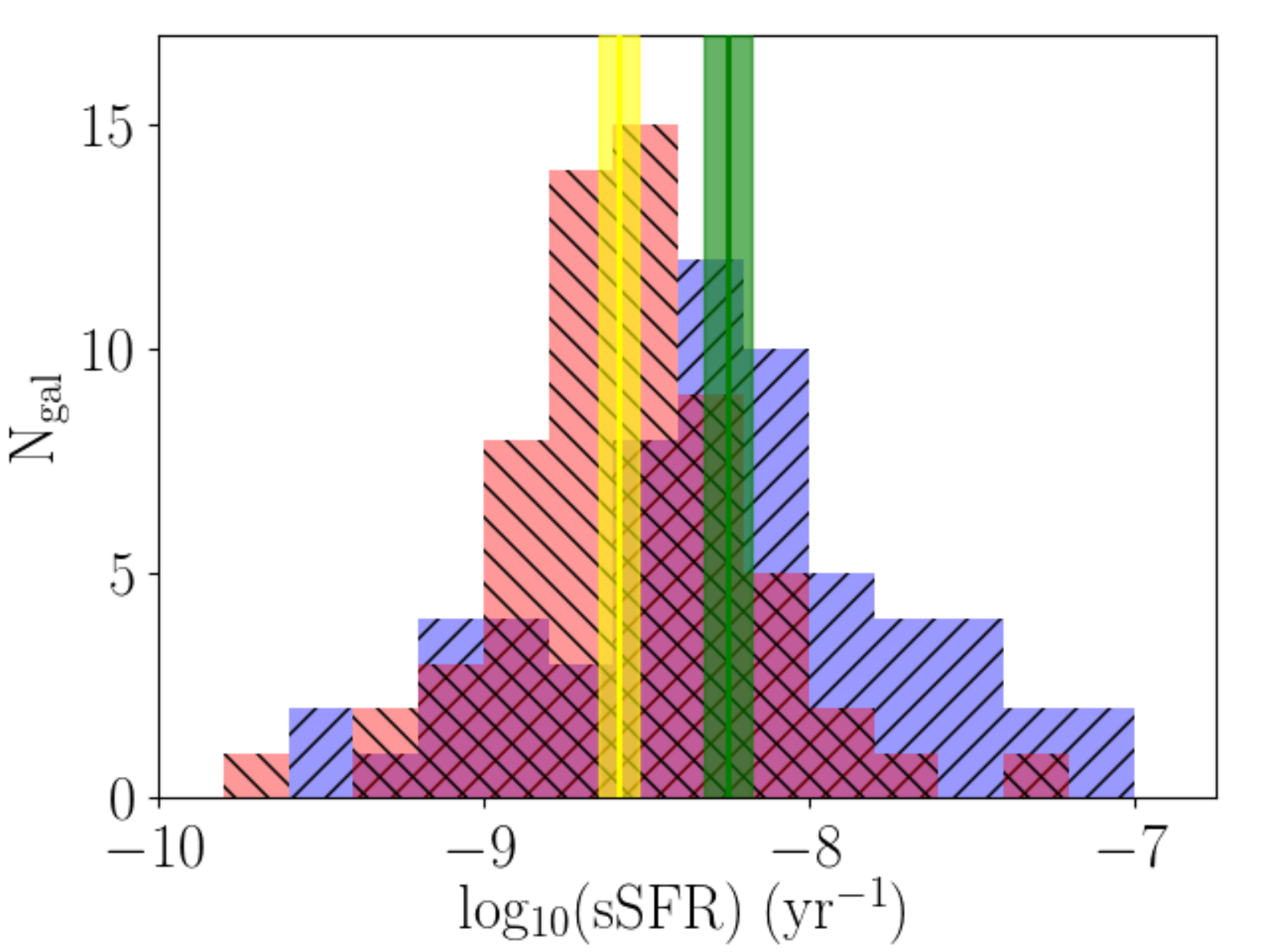}
     }
     \hfill
     \subfloat[]{
       \includegraphics[width=0.32\linewidth]{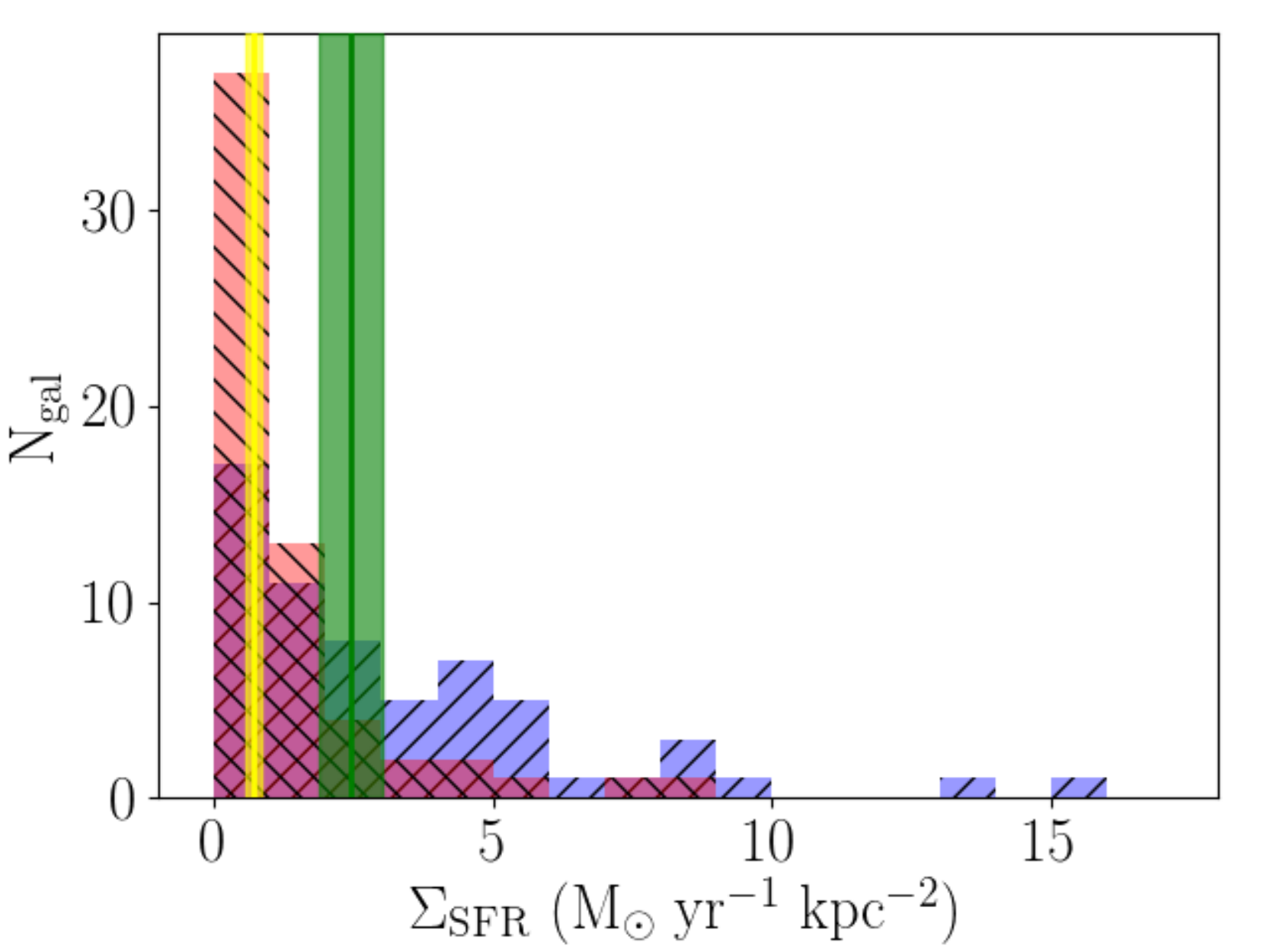}
     }
     \hfill
     \subfloat[]{
       \includegraphics[width=0.32\linewidth]{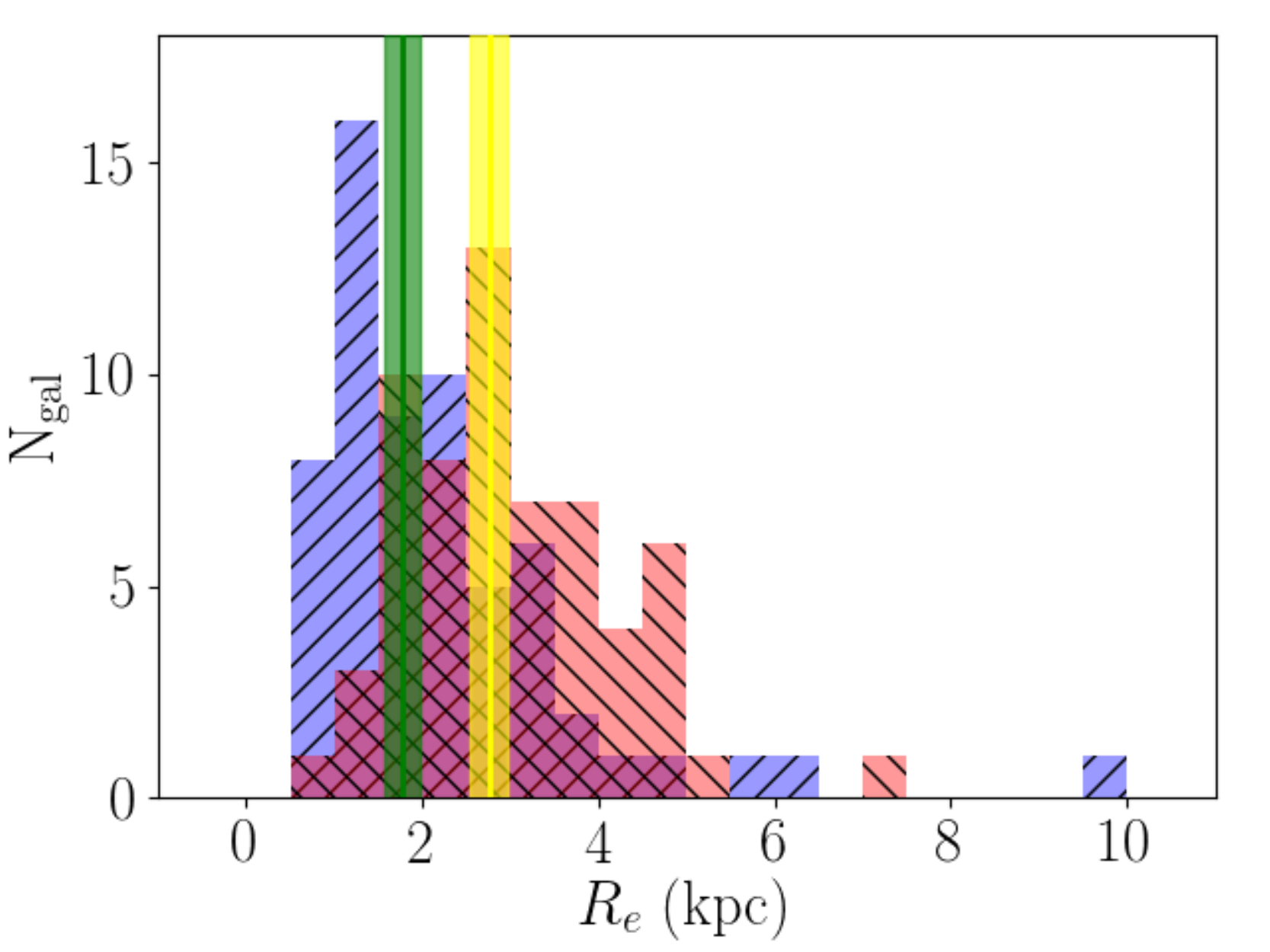}
     }
     \subfloat[]{
       \includegraphics[width=0.32\linewidth]{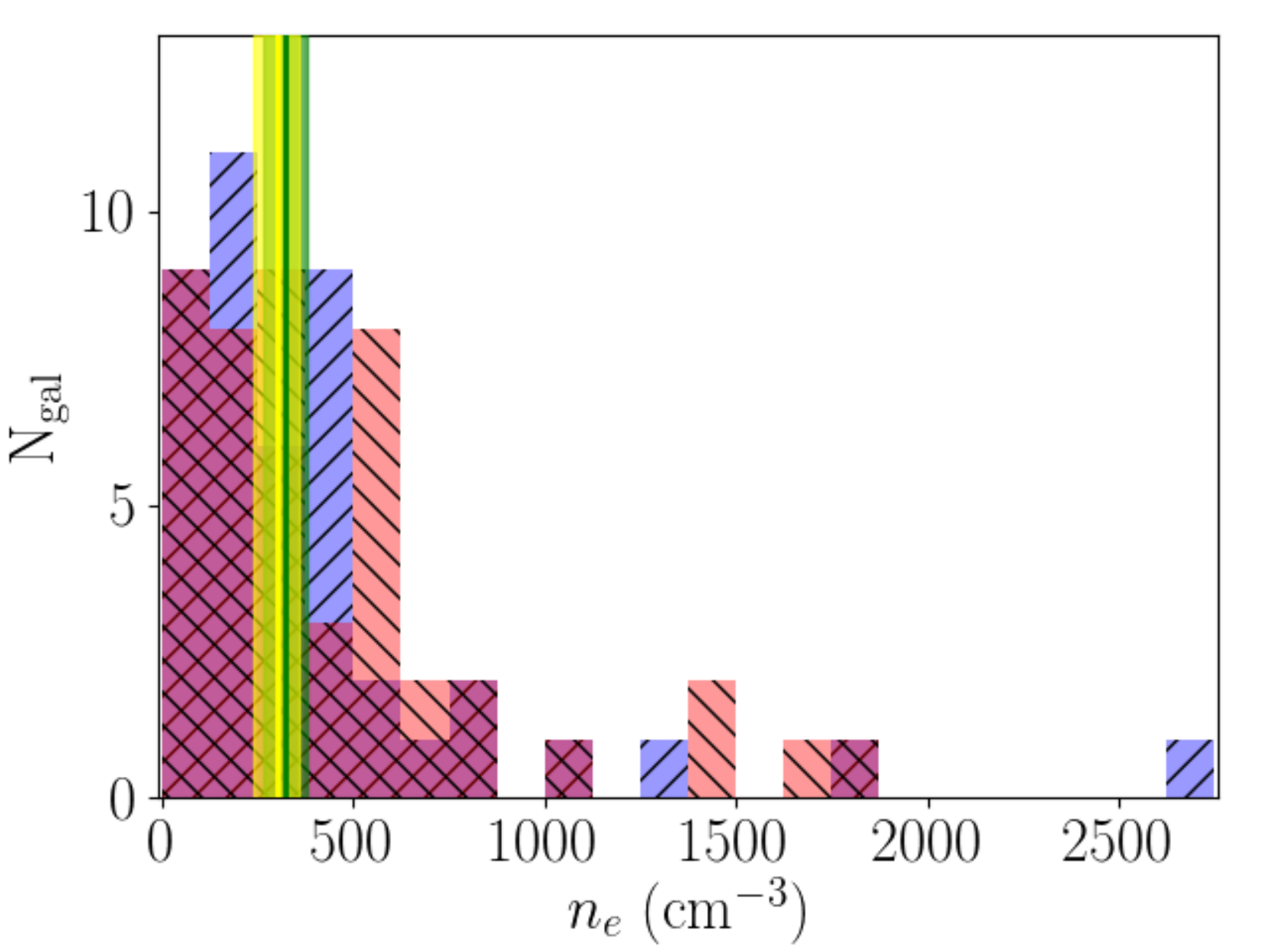}
     }
    \caption{Distribution of physical properties for the 122 galaxy MOSDEF sample shown in Figure \ref{fig:nii_bpt_split} where the blue and red bins correspond to the \textit{high} and \textit{low} samples, respectively.  The green lines are the median values for the \textit{high} population, while the yellow lines are the median values for the \textit{low} population.  The median values with uncertainties for both populations are given in Table \ref{tab:median_high_and_low_samples}.  The following galaxy properties are shown: \textbf{(a)} $M_{\ast}$, \textbf{(b)} log($t/\tau$) of the stellar population using a delayed-$\tau$ star formation model, \textbf{(c)} $t$ of the stellar population assuming a constant star formation history, \textbf{(d)} SFR(H$\alpha$), \textbf{(e)} sSFR, \textbf{(f)} $\Sigma_{\rm{SFR}}$, \textbf{(g)} $R_{\rm{e}}$, and \textbf{(h)} $n_{\rm{e}}$ estimated from the [O~\textsc{II}] or [S~\textsc{II}] doublet emission-lines.  We find that the median $M_{\ast}$, sSFR, $\Sigma_{\rm{SFR}}$, and $R_{\rm{e}}$ differ significantly based on their location in the [N~\textsc{II}] BPT Diagram, while the other properties are the within the uncertainties of each other.}
    \label{fig:galaxy_parameter_split}
\end{figure*}

\subsubsection{Physical Properties of the Galaxies} \label{subsec:galaxy_properties_split}

The fact that there is a segregation in all three plots in Figure \ref{fig:nii_bpt_split} implies that there are some key differences between galaxies offset from the local SDSS sequence and galaxies that overlap with it.  We now investigate several galaxy physical properties to uncover any differences between the galaxies that fall into these two categories.  In this study, as described in Section \ref{subsec:mosdef_sample}, we focus on: $M_{\ast}$, log($t/\tau$) of the stellar population inferred from a delayed-$\tau$ star formation model, $t$ of the stellar population assuming a constant star formation history, SFR(H$\alpha$), sSFR, $\Sigma_{\rm{SFR}}$, $R_{\rm{e}}$, and $n_{\rm{e}}$.  We have measurements of these properties for all 122 galaxies in the sample except for $n_{\rm{e}}$.  As discussed in Section \ref{subsec:mosdef_sample}, a more stringent requirement of a S/N $\geq$ 3 for each component of the [O~\textsc{II}] or [S~\textsc{II}] doublets is needed to estimate $n_{\rm{e}}$ while a S/N $\geq$ 3 for the combined doublet is needed to be plotted on the [S~\textsc{II}] BPT and O$_{32}$ vs. R$_{23}$ diagrams.  Accordingly, while our fiducial sample for analysis contains 122 galaxies, only 90 of them (44 \textit{high} and 46 \textit{low}) have reliable $n_{\rm{e}}$ estimates.

We show how the \textit{high} and \textit{low} samples divide in the space of each of these parameters using histograms, marking the median value of the parameter with solid lines for each \textit{high} and \textit{low} population (Figure \ref{fig:galaxy_parameter_split}).  Showing the data in this format clearly highlights how the \textit{high} and \textit{low} samples separately distribute in each of the galaxy parameters and, accordingly, which of the galaxy parameters are correlated with the location of galaxies on the [N~\textsc{II}] BPT diagram.  The median values for the \textit{high} and \textit{low} populations are listed with 1$\sigma$ uncertainties derived from bootstrap resampling in Table \ref{tab:median_high_and_low_samples}. We also list the probabilities, based on the Kolmogorov-Smirnov (K-S) test, of the null hypothesis that the \textit{high} and \textit{low} samples are drawn from the same parent distribution.

While the distributions of the \textit{high} and \textit{low} samples significantly overlap for some galaxy parameters, there are several parameters for which the \textit{high} and \textit{low} distributions are measurably offset in the median and spread.  These results indicate that \textit{high} sample of MOSDEF galaxies tend to have a smaller $R_{\rm{e}}$ and $M_{\ast}$, but a larger sSFR and $\Sigma_{\rm{SFR}}$ compared to the \textit{low} sample.  We also note that even though $R_{\rm{e}}$, sSFR, and $\Sigma_{\rm{SFR}}$ are correlated with $M_{\ast}$, at fixed $M_{\ast}$ the \textit{high} sample is still found to be smaller with larger sSFR and $\Sigma_{\rm{SFR}}$ values compared to the \textit{low} sample.  The variation in median $\Sigma_{\rm{SFR}}$ between the two samples can be attributed to the difference in median $R_{\rm{e}}$. At fixed size, the median SFR(H$\alpha$) values for the \textit{high} and \textit{low} populations agree within the uncertainties.  With smaller significance, the \textit{high} sample is also younger (smaller median $t/\tau$ and CSF $t$).  We find no mutual correlation for either $n_{\rm{e}}$ or SFR with the the location of galaxies on the [N~\textsc{II}] BPT diagram.  

\begin{table*}
    \centering
    \begin{tabular}{rrrrr}
        \multicolumn{5}{c}{Median Values for Physical Properties of the $High$ and $Low$ Samples} \\
        \hline\hline
        Physical Property & $High$ Median & $Low$ Median & p-value & Statistical significance ($\sigma$) \\
        (1) & (2) & (3) & (4) & (5) \\
        \hline
 log$_{10}$(M$_{\ast}$/M$_{\odot}$) & 9.84 $\pm$ 0.07 & 10.19 $\pm$ 0.06 & 0.0040 & 2.9$\sigma$ \\
 log$_{10}$($t/\tau$) & 0.20 $\pm$ 0.16 & 0.40 $\pm$ 0.08 & 0.2480 & 1.2$\sigma$ \\
 CSF log$_{10}$($t$/yr) & 8.50 $\pm$ 0.08 & 8.60 $\pm$ 0.07 & 0.1671 & 1.4$\sigma$ \\
 SFR(H$\alpha$) (M$_{\odot}$ yr$^{-1}$) & 42.7 $\pm$ 7.5 & 34.3 $\pm$ 4.9 & 0.7923 & 0.3$\sigma$ \\
 sSFR (yr$^{-1}$) & $-$8.25 $\pm$ 0.07 & $-$8.58 $\pm$ 0.06 & 0.0002 & 3.7$\sigma$ \\
 $\Sigma_{\rm{SFR}}$ (M$_{\odot}$ yr$^{-1}$ kpc$^{-2}$) & 2.45 $\pm$ 0.56 & 0.70 $\pm$ 0.15 & 0.0002 & 3.7$\sigma$ \\
 $R_{\rm{e}}$ (kpc) & 1.77 $\pm$ 0.20 & 2.75 $\pm$ 0.21 & 0.00004 & 4.1$\sigma$ \\
 $n_{\rm{e}}$ (cm$^{-3}$) & 322 $\pm$ 58 & 300 $\pm$ 59 & 0.5341 & 0.6$\sigma$ \\
        \hline
    \end{tabular}
    \caption{
  Col. (1): Physical property of the galaxies in the sample shown in Figure \ref{fig:galaxy_parameter_split}.
  Col. (2): Median value with uncertainty of the \textit{high} sample.
  Col. (3): Median value with uncertainty of the \textit{low} sample.
  Col. (4): Two-tailed p-value, based on the K-S test, estimating the probability that the null hypothesis can be rejected.
  Col. (5): Statistical significance (i.e. the $\sigma$ value) that the p-value corresponds to. }
    \label{tab:median_high_and_low_samples}
\end{table*}

\section{Discussion} \label{sec:discussion}

\subsection{Comparison with Previous Work} \label{subsec:comparison_with_previous_work}

We have divided the $z\sim 2.3$ MOSDEF sample according to location in the [N~\textsc{II}] BPT diagram. When separated in this manner, our sample also shows segregation in the [S~\textsc{II}] BPT and O$_{32}$ vs. R$_{23}$ diagrams.  These results update earlier MOSDEF work \citep{sha15, san16}, and are in agreement with more recent MOSDEF studies (e.g., \citealt{sha19, top20a}).  Our findings for the population segregation on the [S~\textsc{II}] BPT diagram are also in agreement with those of \citet{str17}, in which a similar segregation in the [S~\textsc{II}] BPT diagram is found for $z\sim 2.3$ star-forming galaxies from the KBSS survey when separated by [N~\textsc{II}] BPT location.  \citet{str17} did not consider the relative positions of the KBSS equivalent of our \textit{high} and \textit{low} galaxies on the O$_{32}$ vs. R$_{23}$ diagram.  However, these authors do show how photoionization models with different input parameters (i.e., varying ionizing spectrum at fixed nebular metallicity) vary across the O$_{32}$ vs. R$_{23}$ parameter space.  Specifically, it is shown that models with harder ionizing spectra at fixed nebular metallicity are offset towards higher R$_{23}$ at fixed O$_{32}$.  \citet{str17} additionally show that models with harder ionizing spectra produce higher O3 at both fixed S2 and N2.  These separations are at least qualitatively similar to that observed between our \textit{high} and \textit{low} samples. 

The suggested implication of these segregations on both [S~\textsc{II}] BPT and O$_{32}$ vs. R$_{23}$ diagrams is a harder ionizing spectrum at fixed O/H, relative to local galaxies (e.g., \citealt{str17, sha19, top20a}).  A harder ionizing spectrum at fixed nebular metallicity in $z\sim 2.3$ galaxies may arise due to $\alpha$-enhancement (i.e., super-solar O/Fe values) in the massive stars exciting the ionized gas in star-forming regions.  Such abundance patterns may arise naturally in high-redshift galaxies, given the young median ages of their stellar populations (e.g., \citealt{ste16, san20b, top20a}).

We have also shown in Figure \ref{fig:galaxy_parameter_split} that the \textit{high} sample is associated with more compact (i.e., smaller $R_{\rm{e}}$ and higher $\Sigma_{\rm{SFR}}$), and intense (i.e., higher sSFR) star formation than the \textit{low} sample.  These differences between the two populations exist at fixed stellar mass as well.  It will be important to explore the links between these global galaxy properties and the abundance patterns of massive stars using realistic galaxy formation simulations (e.g., FIRE-2; \citealt{hop18}).

\subsection{Comparison with Photoionization Models} \label{subsec:model_comparison}

\subsubsection{Modeling Methodology} \label{subsec:model_methodology}

We use photoionization models to explain simultaneously the joint distributions of \textit{high} and \textit{low} galaxies in the [N~\textsc{II}] and [S~\textsc{II}] BPT and O$_{32}$ vs. R$_{23}$ diagrams.  For this analysis, we use a combination of the code Cloudy (v17.01; \citealt{fer17}) and the Binary Population And Spectral Synthesis (BPASS) v2.2.1 models \citep{eld17, sta18}.  BPASS generates the spectra of model stellar populations, which we used as the input ionizing spectrum to Cloudy.  We assume a constant star-formation history in the BPASS models, as constructed in \citet{top20a}.  Additionally, we use stellar population models that follow a \citet{cha03} IMF and set 100 $M_{\odot}$ as the high-mass cutoff.

Using the BPASS input ionizing spectrum, Cloudy predicts the rest-optical emission-line strengths for different combinations of physical properties.  For this analysis, we set the N/O abundance ratio using equation (2) in \citet{pil12}, use the \citet{asp09} [S~\textsc{II}] abundance pattern, and assume that $n_{\rm{e}} = 250$ cm$^{-3}$, which is characteristic of $z\sim 2.3$ galaxies \citep{san16, str17, har20}.  This value for $n_{\rm{e}}$ is slightly lower than the median values of our \textit{high} and \textit{low} samples; however, it is within the uncertainties of the \textit{low} population and not significantly less than the lower limit of the \textit{high} population median.  The best-fit $n_{\rm{e}}$ values from this study also agree within the uncertainties with the those reported in \citet{san16} and \citet{str17}.  In addition, we use Cloudy to estimate the contribution from the nebular continuum and added this into the BPASS stellar population models.  To estimate this nebular contribution, we assume that log$(U) = -2.5$ and log$(Z_{\rm{neb}}/Z_{\odot}) = -0.2$, which are typical of $z\sim 2.3$ galaxies \citep{san16}.  Changing these conditions does not significantly affect the nebular continuum, which is small contribution to the overall spectrum.  Therefore, adjustments to the initial parameters for the nebular continuum have minimal impact on the final model fitting \citep{top20a}.  As a sanity check on our models, we are able to reproduce the results from \citet{ste16} using their initial input parameters.

For our Cloudy+BPASS model grids, we varied the following parameters: stellar metallicity ($Z_{\ast}$), $t$ (i.e., age) of the stellar population assuming a constant star formation history, gas-phase oxygen abundance ($Z_{\rm{neb}}$), and ionization parameter ($U$).  We find that our choice of $t$ over the range 10$^7$ to 10$^{9.8}$ yr \citep{top20a} has a negligible effect of the position of models on all three diagrams, so we chose a value of 10$^{8.6}$ yr because it is approximately the median $t$ (assuming a constant star-formation history) of the $z\sim 2.3$ MOSDEF sample in this study.  

In the photoionization model grids, ionization parameter and nebular metallicity range between $-3.60$ $\leq$ log($U$) $\leq$ $-1.40$ and $-1.3 \leq \rm{log}(Z_{\rm{neb}}/Z_{\odot}) \leq 0.20$, respectively.  Within the ranges of interest ($-3.40$ $\leq$ log($U$) $\leq$ $-2.40$ and $-1.00 \leq \rm{log}(Z_{\rm{neb}}/Z_{\odot}) \leq 0.00$), these grids are finely sampled in steps of 0.02 dex.  Outside of these regions, log($U$) changes in steps of 0.20 dex. We also include three additional values for $\rm{log}(Z_{\rm{neb}}/Z_{\odot})$: 0.10 and 0.20, extending up towards higher metallicity, and $-1.30$ to round out the low-metallicity extreme.  The stellar metallicity of the BPASS models includes discrete values of $Z_{\ast}$ = 0.00001, 0.0001, 0.001, 0.002, 0.003, 0.004, 0.006, 0.008, 0.01, 0.014, 0.02, and 0.03.  We assume \citet{asp09} solar abundance (i.e., $Z_{\ast}$ = 0.014).

\begin{figure*}
    \centering
    \includegraphics[width=\linewidth]{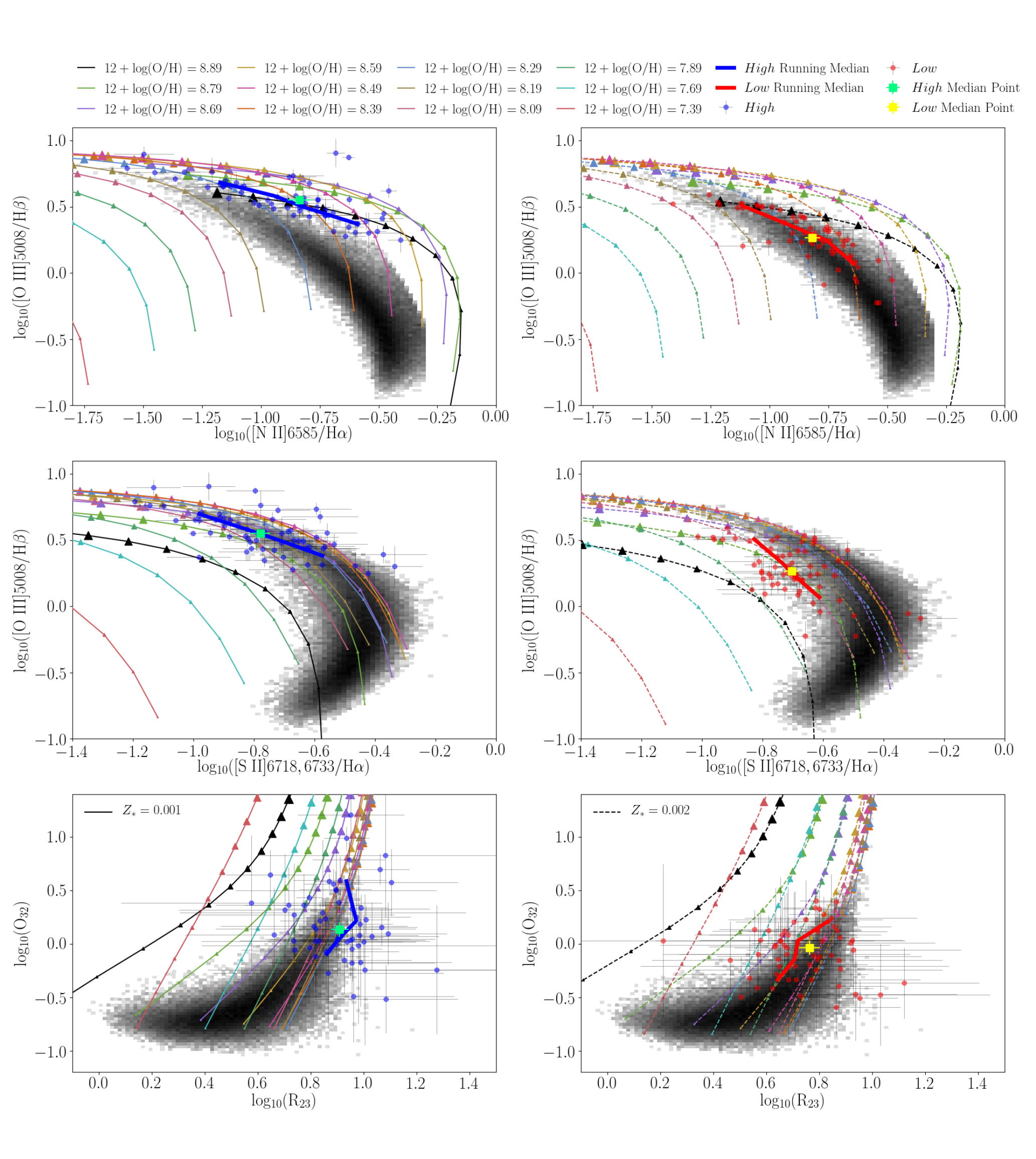}
    \caption{Left column: $Z_{\ast}$ = 0.001 (solid lines).  Right column: $Z_{\ast}$ = 0.002 (dashed lines).  Cloudy+BPASS model emission-line ratios shown on the [N~\textsc{II}] BPT (top row), [S~\textsc{II}] BPT (middle row), and O$_{32}$ vs. R$_{23}$ diagrams (bottom row).  The triangle data points on the curves increase in size as log($U$) increases.  To avoid overcrowding, we vary log($U$) in steps of 0.20 dex and have nebular metallicity range between $-0.60 \leq \rm{12+log(O/H)} \leq 0.20$ in 0.10 dex steps, with additional values that extend deeper into the subsolar regime included as well ($-$0.80, $-$1.00, and $-$1.30).  We adopt an age of 10$^{8.6}$ yr because it is the approximate median age of the MOSDEF sample (given a constant star-formation history), but the results are not sensitive to age.  Also plotted are the local SDSS sample (grey 2D histogram) and the \textit{high} (blue points) and \textit{low} (red points) MOSDEF samples with binned medians as shown in Figure \ref{fig:nii_bpt_split}.  The green and yellow squares, with associated uncertainties, are the median values of the \textit{high} and \textit{low} populations.} \label{fig:photoionization_models}
\end{figure*}

\begin{table*}
    \centering
    \begin{tabular}{rrrr}
        \multicolumn{4}{c}{Single $Z_{\ast}$ Cloudy+BPASS Median Values for Physical Properties of the \textit{High} and \textit{Low} Samples} \\
        \hline\hline
        & Physical Property & \textit{High} Median & \textit{Low} Median \\
        & (1) & (2) & (3) \\
        \hline
Median Points & $Z_{\ast}$ & 0.001 & 0.002 \\
 \\
   & 12+log(O/H) & 8.37$^{+0.04}_{-0.04}$ & 8.31$^{+0.04}_{-0.02}$ \\
 \\ 
   & log($U$) & $-2.88^{+0.04}_{-0.04}$ & $-3.08^{+0.02}_{-0.04}$ \\
        \hline
Large-N2 & $Z_{\ast}$ & 0.002 & 0.003 \\
 \\
   & 12+log(O/H) & 8.49$^{+0.02}_{-0.04}$ & 8.41$^{+0.02}_{-0.02}$ \\
 \\ 
   & log($U$) & $-3.04^{+0.04}_{-0.04}$ & $-3.26^{+0.04}_{-0.04}$ \\
        \hline
Small-N2 & $Z_{\ast}$ & 0.001 & 0.001 \\ 
 \\
   & 12+log(O/H) & 8.31$^{+0.04}_{-0.04}$  & 8.23$^{+0.02}_{-0.02}$ \\
 \\ 
   & log($U$) & $-2.60^{+0.06}_{-0.06}$ & $-2.98^{+0.02}_{-0.02}$ \\
        \hline
    \end{tabular}
    \caption{
  Col. (1): Physical property of the sample.
  Col. (2): Median value with uncertainty of the \textit{high} sample.
  Col. (3): Median value with uncertainty of the \textit{low} sample.  For Columns (2) \& (3), the $Z_{\ast}$ values were selected and were not obtained from fitting the data.  Also, the more finely spaced grids that vary 12+log(O/H) and log($U$) in steps of 0.02 dex were used.}
    \label{tab:single_Zstar_model_properties}
\end{table*}

\begin{figure*}
    \centering
     \subfloat[Median data points]{
       \includegraphics[width=0.49\linewidth]{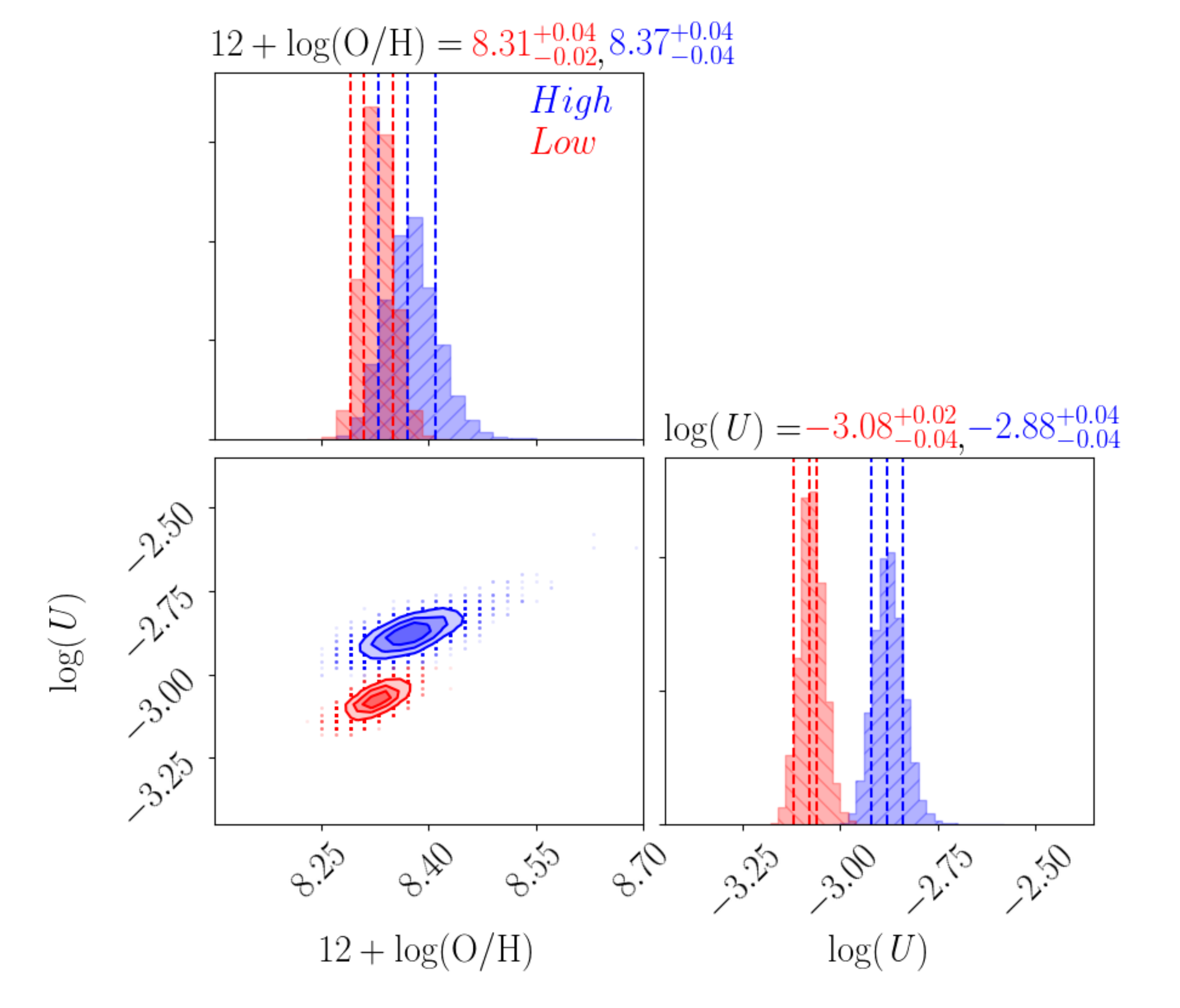}
     }
     \hfill
     \subfloat[Large-N2 data points]{
       \includegraphics[width=0.49\linewidth]{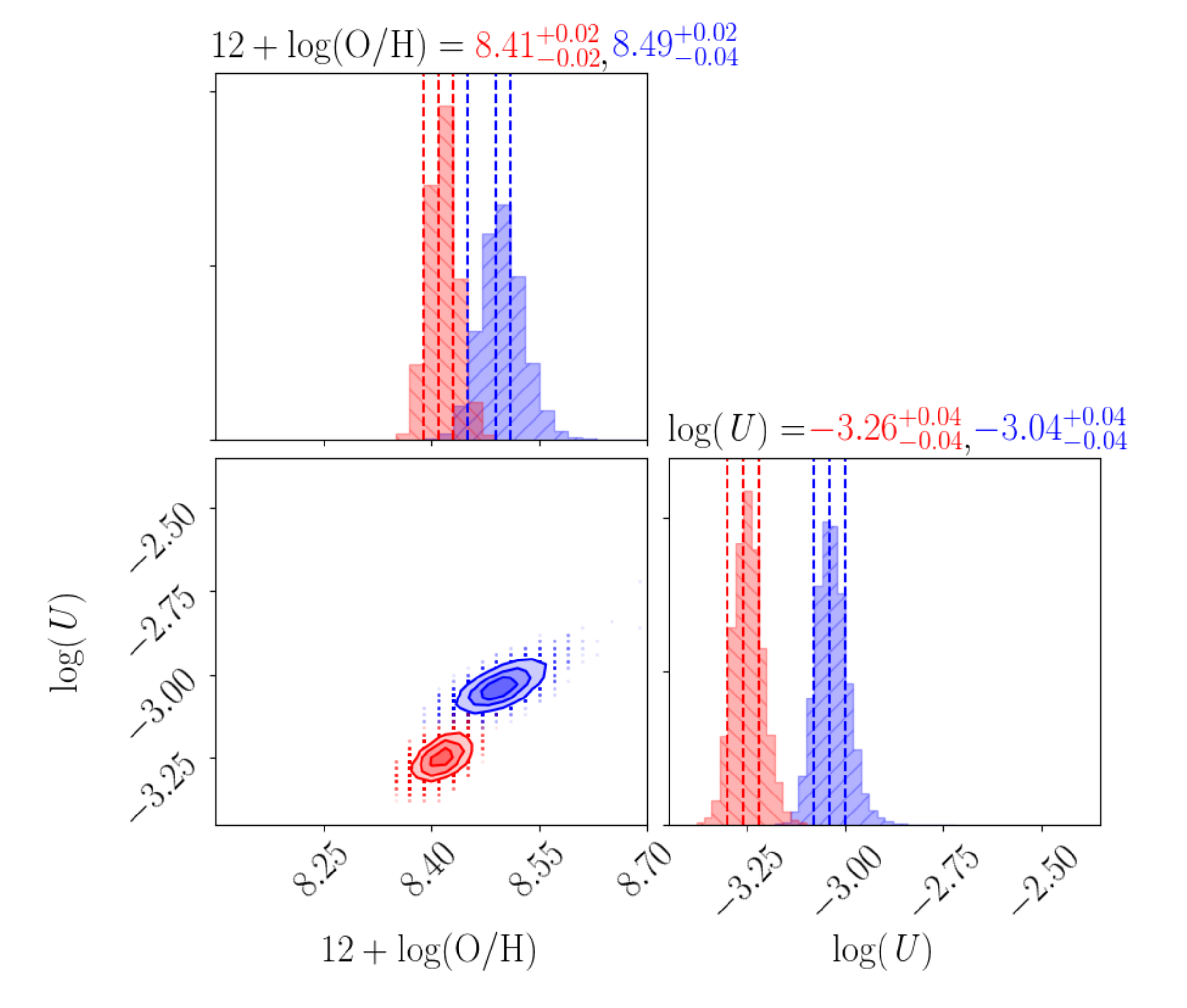}
     }
     \hfill
     \subfloat[Small-N2 data points]{
       \includegraphics[width=0.49\linewidth]{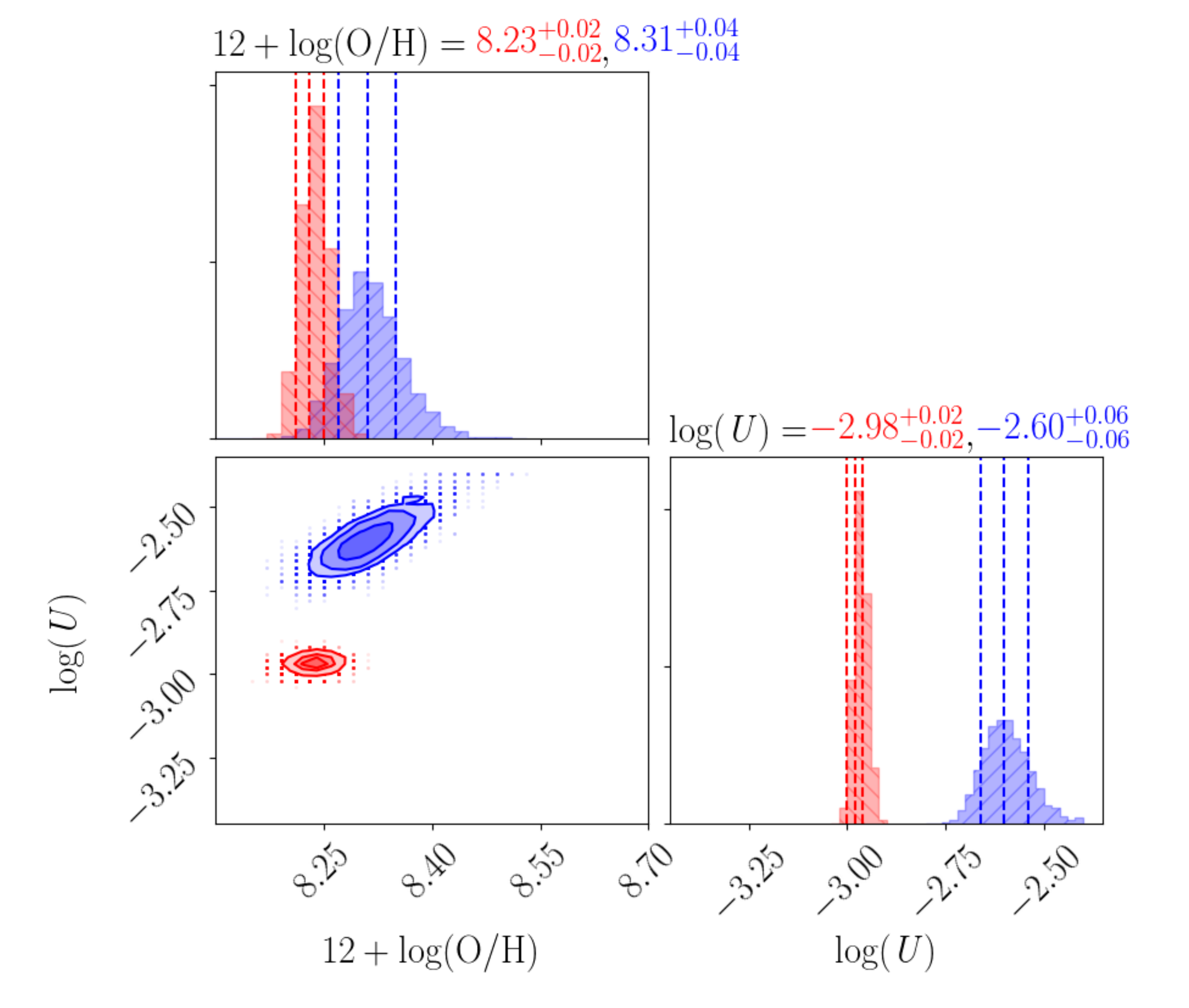}
     }
    \caption{Corner plots comparing the distributions of 12+log(O/H) and log($U$) for the \textit{high} (blue) and \textit{low} (red) MOSDEF populations.  The distributions for the sample median points, i.e. green and yellow squares in Figure \ref{fig:photoionization_models}, (upper left), large-N2 endpoint of the sample binned medians (top right) and small-N2 endpoint of the sample binned medians (bottom center) are included.  We assume $Z_{\ast}$ values of 0.001 and 0.002 (median points; top left panel), 0.002 and 0.003 (large-N2 points; top right panel), and 0.001 and 0.001 (low-N2 points; bottom panel) for the \textit{high} and \textit{low} populations, respectively.  The dashed lines on the histograms mark the median value and 1$\sigma$ uncertainties (16$^{\rm{th}}$ and 84$^{\rm{th}}$ percentiles) for the distributions of 12+log(O/H) and log($U$), and the three contours on the diagonal mark the 1, 1.5, and 2$\sigma$ regions in the 2-D 12+log(O/H)-log($U$) parameter space.  It is shown that moving from large- to small-N2 corresponds to decreasing nebular metallicity and increasing ionization parameter.  In each of the three distributions, the \textit{high} population has a higher 12+log(O/H) and log($U$).} \label{fig:corner_plots}
\end{figure*}

\subsubsection{Fixed $Z_{\ast}$ Models} \label{subsec:single_Zstar_models}

In a recent MOSDEF study, \citet{top20a} combined rest-UV and rest-optical spectra for a sample of 62 star-forming galaxies at $z\sim 2.3$ and constructed \textit{high} and \textit{low} samples using a similar methodology as this study (i.e. based on offset from the local sequence on the [N~\textsc{II}] BPT diagram).  Fitting composite rest-frame UV spectra from their \textit{high} and \textit{low} stacked spectra resulted in best-fit $Z_{\ast}$ values of 0.001 and 0.002, respectively.  Best-fit $Z_{\rm{neb}}$ and log($U$) values were then based on comparing the emission-line ratios measured from the stacked spectra with those generated by the same Cloudy photoionization model described above assuming the stellar metallicities obtained from the rest-UV spectra.  We do not have rest-UV spectra for the full sample of 122 galaxies in our emission-line analysis, and therefore cannot obtain tight constraints on the median $Z_{\ast}$ for the \textit{high} and \textit{low} samples of this work from our rest-optical data alone.  Accordingly, we adopt the best-fit $Z_{\ast}$ values from \citet{top20a} for our two MOSDEF populations.  This is a reasonable assumption because 45/62 (73\%) of the galaxies in \citet{top20a} are in this work, and the \textit{high} and \textit{low} samples are reasonably well matched to the corresponding samples in \citet{top20a}.

Figure \ref{fig:photoionization_models} shows Cloudy+BPASS photoionization models overplotted with our data, assuming $Z_{\ast}$ values of 0.001 (left column; \textit{high} sample) and 0.002 (right column; \textit{low} sample).  Using the solar value of 12+log(O/H) = 8.69 \citep{asp09}, we convert log($Z_{\rm{neb}}/Z_{\odot}$) to 12+log(O/H).  Each curve corresponds to a different 12+log(O/H) value, as indicated by color.  The triangular points along each nebular metallicity curve comprise the sequence of log($U$) values. The points are smallest for log($U$) = $-$3.60 and largest for $-$1.40, therefore, increasing in size as log($U$) increases.  The Cloudy+BPASS model grids are overplotted on the MOSDEF and SDSS data.  The green and yellow squares on the [N~\textsc{II}] BPT, [S~\textsc{II}] BPT, and O$_{32}$ vs. R$_{23}$ diagrams indicate, respectively, the median values of the \textit{high} and \textit{low} samples.  Uncertainties for these median values were obtained through bootstrap resampling and perturbing each resampled data point by its individual error according to a normal distribution.  The sample median was calculated for each bootstrapped, perturbed sample, and then the 16$^{\rm{th}}$ and 84$^{\rm{th}}$ percentiles of the distribution of medians were taken as the lower and upper bounds, respectively, as the $1\sigma$ confidence interval on the plotted median value.

Using stellar metallicities of 0.001 and 0.002, respectively, for the \textit{high} and \textit{low} populations, we then estimate the best-fit 12+log(O/H) and log($U$) values for the sample median data points using a $\chi^2$ method that compares the N2, S2, O3, and O$_{32}$ emission-line ratios between the models and data.  R$_{23}$ is not used because it is not independent of the other emission-line ratios (i.e., R$_{23}$ can be found from a combination of O3 and O$_{32}$).  While Figure \ref{fig:photoionization_models} displays 12+log(O/H) and log($U$) in large steps, we use the finer grids of 0.02 dex spacing for both physical properties when estimating the best-fit sample values.  The results for the best-fit 12+log(O/H) and log($U$) values with 1$\sigma$ uncertainties from the bootstrap resampling are shown in Table \ref{tab:single_Zstar_model_properties}.  The corner plot showing the distributions of these two properties obtained using the bootstrapped, perturbed N2, S2, O3, and O32 medians are shown in the top left panel of Figure \ref{fig:corner_plots}.

We find that both 12+log(O/H) and log($U$) are larger for the median data point of the \textit{high} population (12+log(O/H) = 8.37$^{+0.04}_{-0.04}$, log($U$) = $-2.88^{+0.04}_{-0.04}$) compared to the median data point of the \textit{low} population (12+log(O/H) = 8.31$^{+0.04}_{-0.02}$, log($U$) = $-3.08^{+0.02}_{-0.04}$).  The difference in log($U$) between the two samples is more statistically significant than the difference in 12+log(O/H), as median values of the latter agree within the uncertainties.  Because the median electron densities of the \textit{high} and \textit{low} sample agree within the uncertainties (see Table \ref{tab:median_high_and_low_samples}), the difference in log($U$) can be attributed to the \textit{high} population having a higher number density of ionizing photons.  This is a reasonable assumption given that the \textit{high} sample has a lower median stellar metallicity (i.e., a harder median ionizing spectrum and greater ionizing photon production efficiency) in addition to more concentrated star formation (i.e., higher $\Sigma_{\rm{SFR}}$) compared to the \textit{low} sample.  The \textit{high} population also appears to be more $\alpha$-enhanced, due to it having a lower $Z_{\ast}$ but a higher 12+log(O/H) compared to the \textit{low} population.  Based on the variation of best-fit 12+log(O/H) and log($U$) values between the \textit{high} and \textit{low} median data points, we conclude that quantifying both of these physical parameters, in addition to stellar metallicity (i.e., the hardenss of the ionizing spectrum), is important when explaining the distribution of high-redshift galaxies in the [N~\textsc{II}] BPT diagram.  It is worth noting that we have shown the \textit{high} population to be less massive (i.e., smaller $M_{\ast}$) but more metal-rich (i.e., larger 12+log(O/H)) compared to the \textit{low} population.  This combination of physical properties is the opposite of what is expected based on results from previous studies investigating the mass-metallicity relationship (e.g. \citealt{tre04, ste14, san18, san20b}), which show that $M_{\ast}$ and 12+log(O/H) have a positive relationship.  However, the differences in mass and metallicity between the \textit{high} and \textit{low} samples reflect the scatter in the mass-metallicity relationship.

To check the validity of our assumption that the N/O ratio is consistent between the \textit{high} and \textit{low} populations, we estimate log(N/O) for the sample median points based on the [N~\textsc{II}]$\lambda$6585/[O~\textsc{II}]$\lambda\lambda$3727,3730 tracer. We use the calibration from \citet{str18} to convert from log([N~\textsc{II}]$\lambda$6585/[O~\textsc{II}]$\lambda\lambda$3727,3730) to log(N/O), and find that log(N/O) = $-1.13 \pm 0.05$ and $-1.08 \pm 0.04$ for the \textit{high} and \textit{low} samples, respectively. The consistency in $\log(\mbox{N/O})$ between the \textit{high} and \textit{low} samples validates our Cloudy+BPASS model assumptions and rules out N/O variations as the primary driver of the [N~\textsc{II}] BPT offset. This conclusion agrees with results from other current MOSDEF studies (e.g. \citealt{sha19, san20b, san20a, top20b}) as well as other $z\sim 2$ studies (e.g. KBSS; \citealt{ste14, str17, str18}).

For the most part, these results are in agreement with \citet{top20a}.  The best fit 12+log(O/H) and log($U$) values from both studies are in agreement within 1$\sigma$, except for the log($U$) of the two \textit{high} populations, which agree within 2$\sigma$.  However, this study finds a larger difference in log($U$) between the \textit{high} and \textit{low} samples, 0.20 dex, compared to \citet{top20a}, 0.07 dex.  On the other hand, \citet{top20a} find a larger variance in 12+log(O/H) between the two populations, 0.10 dex, compared to this study, 0.06 dex.  These differences are due to the slightly different median rest-optical line ratios for the \textit{high} and \textit{low} populations in \citet{top20a} and in the current work.

To characterize the variation of physical properties within each of our \textit{high} and \textit{low} populations, we apply the same methodology of finding the best-fit 12+log(O/H) and log($U$) values, but in this case for the two endpoints of the binned medians for the \textit{high} and \textit{low} samples in the [N~\textsc{II}] BPT diagram.  We use the same binning method for the [N~\textsc{II}] BPT diagram (i.e., four equally sized bins based on O3N2 strength; see Section \ref{subsec:nii_bpt_split} above for complete details on binning) when calculating the median O3, N2, S2, and O$_{32}$ emission-line ratios for the endpoints of both binned median lines.  Uncertainties for the line ratios of the median endpoints are estimated using the same methodology of bootstrap resampling coupled with perturbation of the datapoints of individual galaxies according to their error bars, as described above.  On the [N~\textsc{II}] BPT diagram, the binned medians move primarily in the N2 direction, therefore, we refer to the endpoints of both binned median lines as \textquotedblleft large-N2'' and \textquotedblleft small-N2''.  It is important to note that the large-N2 endpoints of both \textit{high} and \textit{low} binned median lines have larger S2, and smaller O$_{32}$ compared to the small-N2 endpoints.  In other words, while the binning method was not the same on all three diagrams in Figure \ref{fig:nii_bpt_split} (i.e., the four bins do not necessarily contain the same galaxies), the large-N2 bins on the [N~\textsc{II}] BPT diagram have significant overlap with (and therefore roughly correspond to) the larger-S2, and smaller-O$_{32}$ endpoints of the binned median lines on the [S~\textsc{II}] BPT and O$_{32}$ vs. R$_{23}$ diagrams.  

The results of \citet{top20b} suggest that a positive correlation exists between N2 and $Z_{\ast}$ (i.e., stellar metallicity tends to be higher at larger N2 values).  Therefore, we vary the $Z_{\ast}$ values that we assume for the large-N2 and small-N2 endpoints of the binned median lines accordingly.  For the \textit{low} population, we assume stellar metallicities of 0.003 and 0.001 (i.e., $\pm$0.001 from the median $Z_{\ast}$) for the large-N2 and small-N2 points, respectively.  Similarly, for the \textit{high} population, we assume stellar metallicity values of 0.002 and 0.001 for the large-N2 and small-N2 points, respectively.  We do not lower the stellar metallicity of the small-N2 point of the \textit{high} population to our next available $Z_{\ast}$ model value (0.0001, a factor of 10 lower than the next highest $Z_{\ast}$ value) because the results from \citet{top20b} do not suggest that such a significant variation in stellar metallicity is observed.  It is important to note that while we do vary $Z_{\ast}$ in accordance with the results from \citet{top20a}, the best-fit 12+log(O/H) and log($U$) for the large-N2 and small-N2 endpoints of the \textit{high} and \textit{low} populations do not change significantly if we use $Z_{\ast}$ = 0.001 and 0.002 (i.e., the $Z_{\ast}$ values used for the median \textit{high} and \textit{low} population data points).  We will discuss this systematic effect of how our choice in $Z_{\ast}$ influences the best-fit 12+log(O/H) and log($U$) in more detail below in Sections \ref{subsec:multi_zstar_models} and \ref{subsec:model_summary}.  

Median 12+log(O/H) and log($U$) values for the large-N2 and small-N2 endpoints of the \textit{high} and \textit{low} populations are included in Table \ref{tab:single_Zstar_model_properties}.  The corner plots showing the distributions of 12+log(O/H) and log($U$) for the large-N2 and small-N2 endpoints are shown in the top right and bottom panels, respectively, of Figure \ref{fig:corner_plots}.  In both samples we find an anti-correlation between 12+log(O/H) and log($U$) along the binned median lines.  Moving from large-N2 to small-N2 (therefore also large-S2 to small-S2 and small-O$_{32}$ to large-O$_{32}$) leads to a lower 12+log(O/H) but a higher log($U$).  This anti-correlation between log($U$) and 12+log(O/H) has been observed in local star-forming galaxies and H II regions \citep{per14}.  In addition, this trend is consistent with the idea that the high O$_{32}$ tail of the local sequence on the O$_{32}$ vs. R$_{23}$ diagram includes galaxies with low metallicity, and high ionization parameter \citep{sha15}.  Also, on the [N~\textsc{II}] BPT diagram both stellar and nebular metallicity decrease along the local sequence (i.e., from low O3 and high N2 to high O3 and low N2).

\begin{figure*}
    \centering
    \includegraphics[width=\linewidth]{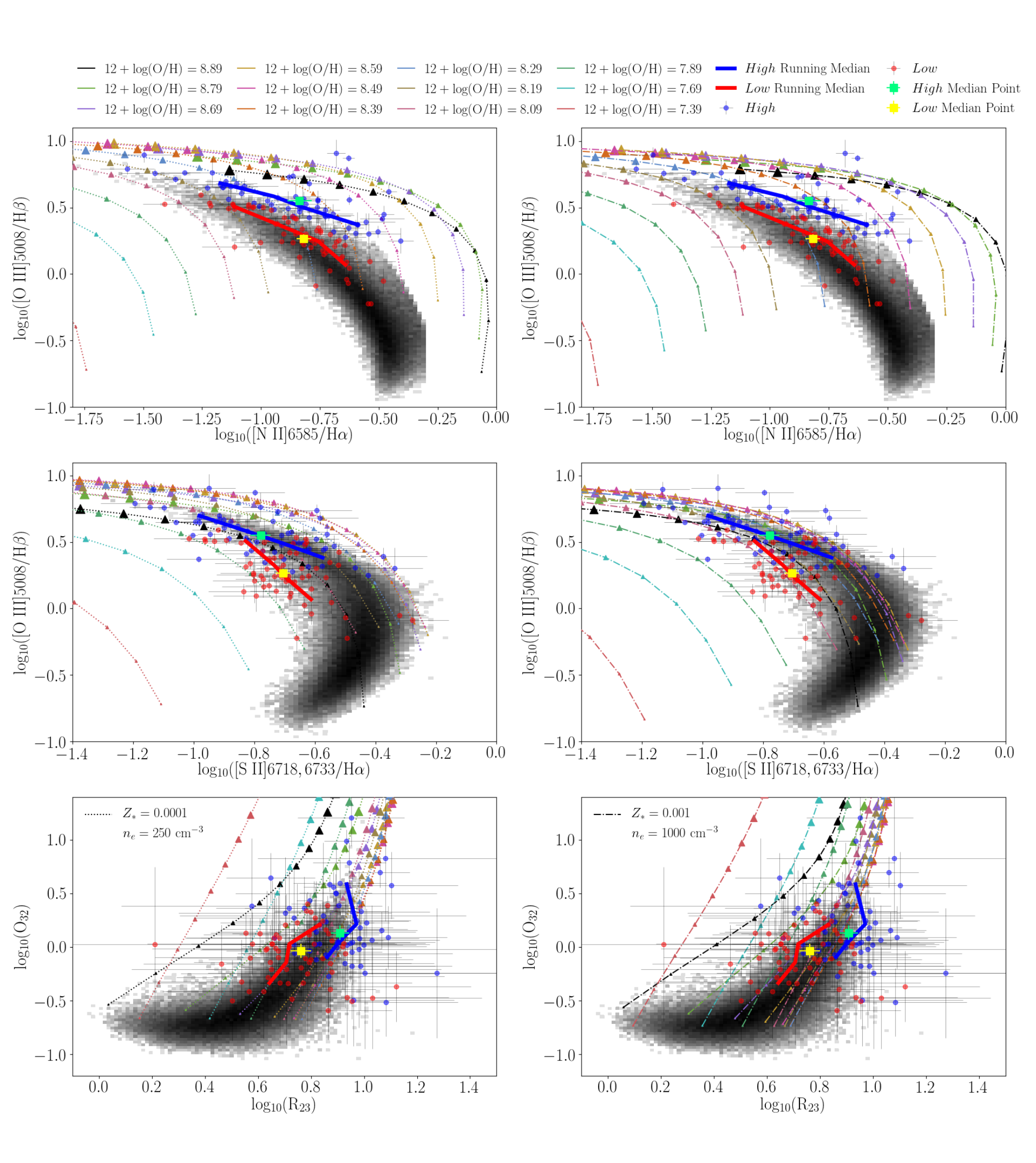}
    \caption{Left column: $Z_{\ast}$ = 0.0001, $n_{\rm{e}} = 250$ cm$^{-3}$ (dotted lines).  Right column: $Z_{\ast}$ = 0.001, $n_{\rm{e}} = 1000$ cm$^{-3}$ (dot-dashed lines).  Cloudy+BPASS model emission-line ratios shown on the [N~\textsc{II}] BPT (top row), [S~\textsc{II}] BPT (middle row), and O$_{32}$ vs. R$_{23}$ diagrams (bottom row).  A comparison of these models with the corresponding panels in Figure \ref{fig:photoionization_models} shows how raising the electron density and lowering stellar metallicity affects the predicted emission-line ratios.  The correspondence between colors and symbol size and, respectively, 12+log(O/H) and log($U$), matches that in Figure \ref{fig:photoionization_models}.  Once again, we adopt an age of 10$^{8.6}$ yr because it is the approximate median age of the MOSDEF sample (given a constant star-formation history) and include the local SDSS sample (grey 2D histogram), the \textit{high} (blue) sample, and \textit{low} (red) sample with binned medians as shown in Figure \ref{fig:nii_bpt_split}.  The green and yellow squares, with associated uncertainties, are the median values of the \textit{high} and \textit{low} populations as shown in Figure \ref{fig:photoionization_models}.  Aside from at roughly solar and supersolar nebular metallicities, a higher electron density has a minimal affect on the Cloudy+BPASS model grids and cannot reach the elevated O3 and O$_{32}$ galaxies.  A lower stellar metallicity is able to reach these values, suggesting that the most offset galaxies have the hardest ionizing spectrum.} \label{fig:photoionization_models_n_e_comparison}
\end{figure*}

\subsubsection{Variable $Z_{\ast}$ Models} \label{subsec:multi_zstar_models}

As described above, thus far we have assumed $Z_{\ast}$ values of 0.001 and 0.002 for the \textit{high} and \textit{low} population median points to match the best-fit values from modeling the rest-frame UV spectra of $z\sim 2.3$ MOSDEF galaxies \citep{top20a}.  While this is a reasonable assumption, we also investigate how relaxing our requirement on stellar metallicity affects the best-fit nebular metallicity and ionization parameter values of the two populations.  For this additional analysis, we use the same $\chi^2$ method described above (i.e., fitting the N2, S2, O3, and O$_{32}$ emission-line ratios). However, we not only fit for 12+log(O/H) and log($U$), but also treat $Z_{\ast}$ as a free parameter.  We investigate two different model grids: one that allows a small range of $Z_{\ast}$ for the \textit{high} and \textit{low} samples, including the best-fit stellar metallicities from \citet{top20a} for each population and those that are adjacent to the best-fit values, and a fully unrestricted method that allows all $Z_{\ast}$ values used by the BPASS models ranging from 0.00001 to 0.03.  

This analysis shows that when we relax the constraints provided by rest-UV spectra (both with limited freedom and in a fully unconstrained manner), the rest-optical emission lines tend to favor higher stellar metallicities.  When we allow for limited freedom in $Z_{\ast}$ the stellar metallicities for the \textit{high} and \textit{low} samples prefer the largest allowed $Z_{\ast}$ $-$ i.e., 0.002 and 0.003, respectively.  When all stellar metallicity values are allowed, the \textit{high} and \textit{low} populations favor $Z_{\ast}$ values about 4-5 times greater than those found by \citet{top20a} based on fitting rest-UV spectra.  These shifts demonstrate the importance of imposing external constraints from a more direct probe of massive stars and stellar metallicity, i.e., the the rest-UV continuum, as opposed to relying only on a joint fit of rest-optical emission lines. We also note that, while the best-fit values of $Z_{\ast}$ increase when stellar metallicity is allowed to vary, the \textit{low} population still always favors a higher $Z_{\ast}$ than the \textit{high} population.  

The best-fit 12+log(O/H) and log($U$) values for the single-$Z_{\ast}$ and limited freedom $Z_{\ast}$ models are significantly consistent (i.e., within 1$\sigma$) for both the \textit{high} and \textit{low} populations.  For the \textit{high} sample, the best-fit values for the single-$Z_{\ast}$ and fully unconstrained $Z_{\ast}$ models are significantly consistent as well.  For the \textit{low} sample, only some of the best-fit values for the single-$Z_{\ast}$ and fully unconstrained $Z_{\ast}$ models are consistent within 1$\sigma$; however, all values are consistent within 2$\sigma$.  In addition, similar to what we find for the single-$Z_{\ast}$ models, we find that O/H increases and log($U$) decreases with increasing N2 in both free-$Z_{\ast}$ models.

The results from this section provide guidance on which wavelength regimes best constrain different galaxy properties. Supported by other studies (e.g. \citealt{top20a}), rest-UV spectra are essential for accurately estimating $Z_{\ast}$. Rest-optical spectra, even with the combination of multiple emission-line ratios, cannot constrain the stellar metallicity.  At high redshift, we have shown that modeling rest-optical spectra alone will lead to overestimates of stellar metallicity relative to more accurate methods that incorporate rest-UV spectral information.  However, rest-optical spectra can be used to constrain 12+log(O/H) and log($U$).  Because the best-fit values of nebular metallicity and ionization parameter vary based on which rest-optical emission-line diagram is used (see Figure \ref{fig:photoionization_models}), the combination of line ratios from multiple diagrams will increase the accuracy of the models. These parameters can even be estimated with reasonable precision without detailed knowledge of the ionizing spectrum (i.e., the metallicity of the stellar population).

\subsubsection{The Importance of $\alpha$-Enhancement} \label{subsec:model_summary}

In summary, using rest-optical emission line ratios alone without the constraints from fitting rest-UV spectra, we infer systematically higher $Z_{\ast}$ values for both \textit{high} and \textit{low} samples.  At the same time, when multiple rest-optical emission lines ratios are measured (N2, S2, O3, and O$_{32}$), the inferred median nebular parameters (12+log(O/H) and log($U$)) and their variation across the BPT diagram, do not depend strongly on the allowed range of $Z_{\ast}$.  However, the inclusion of constraints on $Z_{\ast}$ from rest-UV fitting is essential for our understanding of the abundance patterns (i.e., $\alpha$-enhancement) of $z\sim 2.3$ galaxies and we adopt the constrained values of \citet{top20a} for our fiducial modeling procedure (Table \ref{tab:single_Zstar_model_properties}).  Because we see variation in $Z_{\ast}$, 12+log(O/H), and log($U$) at the median and large- and small-N2 endpoints for the \textit{high} and \textit{low} samples, we conclude that constraints on all three of these physical properties are required for fully understanding the observed distribution of $z\sim 2.3$ star-forming galaxies in the [N~\textsc{II}] BPT diagram. The importance of $\alpha$-enhancement has also been highlighted by \citet{ste16} and \citet{cul19}.

It is also important to note that while the \textit{high} sample is more $\alpha$-enhanced with a harder ionizing spectrum compared to the \textit{low} sample, both populations are $\alpha$-enhanced and both are characterized by a harder ionizing spectra when compared with their equivalents (i.e., galaxies with similar rest-optical line ratios) at low redshift. Notably, such physical differences apply to the \textit{low} population, even though it overlaps with the local [N~\textsc{II}] BPT sequence.  Specifically, an overlap in emission-line ratios does not correspond to the same properties for the ionizing radiation field and ionized gas.  Therefore, we recommend determining the location of galaxies in multiple rest-optical emission-line diagrams, as well as having some knowledge of the ionizing spectrum, when trying to identify local systems that can be used as true $z \sim 2$ analogs.

In addition, the differences between the \textit{high} and \textit{low} populations found in both this section and in Section~\ref{subsec:galaxy_properties_split} suggest that $z\sim 2.3$ galaxies in the \textit{high} sample that generally have a lower $Z_{\ast}$ with higher 12+log(O/H) and log($U$) values are also typically smaller with higher sSFR and $\Sigma_{\rm{SFR}}$ values.  Looking ahead, we must understand why there exists a connection between the local properties of ionized star-forming regions and the overall compactness/intensity of star formation. However, exploring this phenomenon is outside the scope of the current work.

\subsubsection{Discrepancies Between Data and Models} \label{subsec:model_data_discrepancies}

It is worth noting that in all three diagrams in Figure \ref{fig:photoionization_models}, there are galaxies that do not lie within the Cloudy+BPASS grid space.  On the [N~\textsc{II}] BPT and [S~\textsc{II}] BPT diagrams, there are galaxies with O3 values that exceed the extremes of the model grids.  Similarly on the O$_{32}$ vs. R$_{23}$ diagram, we observe MOSDEF galaxies with higher R$_{23}$ values than what the Cloudy+BPASS models predict.  These offsets are primarily found among the \textit{high} population.  Such extreme galaxies are not specific to this study, as \citet{str17} find similar discrepancies between data and models in the corresponding emission-line diagnostic diagrams.

For the Cloudy+BPASS model grids in this study, two possible solutions could be to either elevate $n_{\rm{e}}$ or lower $Z_{\ast}$.  Increasing $n_{\rm{e}}$ has been shown to cause elevated O3 and N2 values \citep{kew13}.  Since the \textit{high} population (defined by its elevated O3 and/or N2 values) also has elevated R$_{23}$ values on average, increasing $n_{\rm{e}}$ in our Cloudy+PASS models theoretically could push the curves to higher R$_{23}$.  The models in this study, which were constructed by \citet{top20a}, set $n_{\rm{e}}$ = 250 cm$^{-3}$, which is typical of $z\sim 2.3$ galaxies in both the KBSS sample \citep{str17} and early MOSDEF work \citep{san16}.  It is also comparable with the median electron densities reported in this work.  The models employed in \citet{str17} use $n_{\rm{e}}$ = 300 cm$^{-3}$, and similarly cannot reach the highest observed O3 and R$_{23}$ values in the KBSS survey.

Using $Z_{\ast}$ = 0.001, we explore raising $n_{\rm{e}}$ to 1000 cm$^{-3}$ in the Cloudy+BPASS models (right column of Figure \ref{fig:photoionization_models_n_e_comparison}).  Comparing the elevated electron density models with the $Z_{\ast}$ = 0.001, $n_{\rm{e}}$ = 250 cm$^{-3}$ scenario (left column of Figure \ref{fig:photoionization_models}) shows the differences in the predicted emission line ratios caused by isolating variations in $n_{\rm{e}}$ on the [N~\textsc{II}] BPT, [S~\textsc{II}] BPT, and O$_{32}$ vs. R$_{23}$ diagrams.  We find that elevating $n_{\rm{e}}$ to such an extreme value has only a small effect on the emission-line ratios, and the models still do not encompass the galaxies observed at the highest N2, S2, O3, and R$_{23}$ values.  

A more promising approach is to lower the stellar metallicity.  The left column of Figure \ref{fig:photoionization_models_n_e_comparison} gives the model grids for $Z_{\ast}$ = 0.0001, which reach the majority of the extreme data points.  Therefore, a very low $Z_{\ast}$ (i.e., a very hard ionizing spectrum) is the more likely than variations in $n_{\rm{e}}$ to be cause for these outliers.  

In addition to a low $Z_{\ast}$, other effects may also lead to a harder ionizing spectrum (e.g., variations in the IMF slope and high-mass cutoff or AGN partially contributing to the integrated emission lines).  
Systematic uncertainties such as the BPASS models under-predicting the hardness of the ionizing spectrum at a given $Z_{\ast}$ (particularly at lower values of $Z_{\ast}$), different star-formation histories, or uncertainties in the dust corrections applied to O$_{32}$ and R$_{23}$ could play a role in the observed discrepancies between models and a minority of the sample as well.

\section{Summary} \label{sec:summary}

We present results on the emission-line properties of a sample of 122 star-forming galaxies at $1.9\leq z\leq 2.7$ from the complete MOSDEF survey with $\geq3\sigma$ detections for the [O~\textsc{II}]$\lambda\lambda$3727,3730, H$\beta$, [O~\textsc{III}]$\lambda\lambda$4960,5008, H$\alpha$, [N~\textsc{II}]$\lambda$6585, and [S~\textsc{II}]$\lambda\lambda$6718,6733 emission lines.  To investigate the observed systematic offset of $z>1$ star-forming galaxies on the [N~\textsc{II}] BPT diagram relative to local systems, the MOSDEF sample is split into the \textit{high} (offset with elevated O3 and/or N2 values) and \textit{low} (overlapping with the local SDSS sequence) samples.  We compare the location of galaxies in both populations on the [S~\textsc{II}] BPT and O$_{32}$ vs. R$_{23}$ diagrams.  Additionally, we compare physical properties $-$ SFR, $M_{\ast}$, $R_{\rm{e}}$, sSFR, $\Sigma_{\rm{SFR}}$, $n_{\rm{e}}$, and $t$ (i.e., stellar population age) $-$ of the galaxies in both groups using additional CANDELS and 3D-\textit{HST} ancillary data.  Finally, we use Cloudy+BPASS photoionization models to investigate potential differences in the stellar metallicity, nebular metallicity, and ionization parameter of the \textit{high} and \textit{low} groups.

The main results are as follows:
\begin{enumerate}
    \item On the [S~\textsc{II}] BPT and O$_{32}$ vs. R$_{23}$ diagrams, the \textit{high} sample is offset on average towards higher S2 at fixed O3 (or vice versa) and higher R$_{23}$ value at fixed O$_{32}$, respectively, relative to the \textit{low} sample.  These results update earlier work from MOSDEF \citep{san16} and are consistent with results from \citet{str17}, based on the KBSS survey.
    \item The \textit{high} sample has a smaller median $R_{\rm{e}}$ and $M_{\ast}$, but higher median sSFR and $\Sigma_{\rm{SFR}}$ compared to the \textit{low} sample.  The observed differences in $R_{\rm{e}}$, sSFR, and $\Sigma_{\rm{SFR}}$ are present at fixed stellar mass.  The \textit{high} sample is also slightly younger in median age ($t$); however this age difference is not significant within the uncertainties.  There is not a significant variation in SFR or $n_{\rm{e}}$ for the two populations.  These results imply that the \textit{high} population is associated with more concentrated and intense star formation than the \textit{low} population.
    \item Using photoionization models, we find that the \textit{high} population has larger 12+log(O/H) and log($U$) values compared to the \textit{low} population.  These conclusions hold both when we use the results from \citet{top20a} and assume $Z_{\ast}$ values of 0.001 and 0.002 for the \textit{high} and \textit{low} samples, respectively, as well as when we treat $Z_{\ast}$ as a free parameter.  Also, while the lower $Z_{\ast}$/$Z_{\rm{neb}}$ ratio of the \textit{high} population implies that this sample is more $\alpha$-enhanced than the \textit{low} population, both samples are significantly $\alpha$-enhanced relative to local star-forming galaxies with similar emission-line ratios assuming typical local galaxies have roughly solar $\alpha$/Fe. These differences in the properties of the ionizing radiation field are critical to include when describing high-redshift galaxies -- even those that overlap the local emission-line sequences.
    \item Combining the results from the median galaxy properties of the \textit{high} and \textit{low} samples with the Cloudy+BPASS model grids leads to the conclusion that galaxies with a harder ionizing spectrum are associated with smaller sizes and higher sSFR and $\Sigma_{\rm{SFR}}$.  
    \item A harder ionizing spectrum at fixed nebular metallicity is favored as one of the key drivers of the [N~\textsc{II}] BPT offset.  In this study, we find that variation in multiple physical parameters drives the variation in emission-lines in the MOSDEF sample, including $Z_{\ast}$, 12+log(O/H), and log($U$). However, even in regions where the $z\sim 2.3$ galaxies overlap local ones in the space of emission-line ratios, the inferred underlying physical parameters for $z\sim 2.3$ galaxies are distinct from those of local galaxies. The full set of these parameters must be understood through modeling both rest-UV and rest-optical spectra, in order to understand the translation between empirical emission-line ratios and key physical quantities such as the gas-phase oxygen abundance.
\end{enumerate}

Understanding the global properties of high-redshift galaxies using large samples at $z\sim 2.3$ is essential when comparing them with local galaxies.  The use of large high-redshift samples has enabled us to discover more subtle trends in the MOSDEF data that were not found with more limited data sets (e.g. \citealt{sha15,san16}).  The work from this study has shown that galaxies with a harder ionizing spectrum also tend to be smaller in size with more intense and compact star formation.  Why these factors are linked is not yet clear, and combining large $z\sim 2.3$ samples of star-forming galaxies with realistic galaxy formation simulations (e.g., FIRE-2; \citealt{hop18}) will be important in finding this connection.  

It has also been shown in both this work and in \citet{str17} that models are not yet able to reach the highest observed O3 and R$_{23}$ values for $z\sim 2.3$ star-forming galaxies on the [N~\textsc{II}] BPT, [S~\textsc{II}] BPT, and O$_{32}$ vs. R$_{23}$ diagrams.  The local sequence does not reach such elevated O3 and R$_{23}$ values, and unlike their high redshift counterparts, do not greatly exceed the data space covered by the model grids.  This discrepancy indicates a need to develop both stellar population and emission-line models that better match the now large data sets of $z\sim 2.3$ star-forming galaxies.  

\section*{Acknowledgements}
We thank the referee for a thorough and constructive report. We acknowledge support from NSF AAG grants AST1312780, 1312547, 1312764, and 1313171, grant AR-13907 from the Space Telescope Science Institute, and grant NNX16AF54G from the NASA ADAP program. We also acknowledge a NASA contract supporting the ``WFIRST Extragalactic Potential Observations (EXPO) Science Investigation Team'' (15-WFIRST15-0004), administered by GSFC.  We thank the 3D-\textit{HST} collaboration, who provided us with spectroscopic and photometric catalogs used to select MOSDEF targets and derive stellar population parameters.  We acknowledge the First Carnegie Symposium in Honor of Leonard Searle for useful information and discussions that benefited this work.  This research made use of Astropy,\footnote[14]{http://www.astropy.org} a community-developed core Python package for Astronomy \citep{ast13, ast18}.  We finally wish to extend special thanks to those of Hawaiian ancestry on whose sacred mountain we are privileged to be guests.

\section*{Data Availability}
The data underlying this article will be shared on reasonable request to the corresponding author.

\vspace{5mm}
\textit{Facilities}: \textit{Keck}/MOSFIRE, \textit{SDSS}

\textit{Software}: Astropy \citep{ast13, ast18}, Corner \citep{for16}, IPython \citep{per07}, Matplotlib \citep{hun07}, NumPy \citep{oli06, van11}, SciPy \citep{oli07, mil11, vir20}

\bibliographystyle{mnras}
\bibliography{references}

\bsp
\label{lastpage}
\end{document}